\documentclass[prl,aps,twocolumn,amsmath,amssymb]{revtex4-1}
\pdfoutput=1
\usepackage{hyperref}
\usepackage{graphicx}
\usepackage{color}
\usepackage{multirow}

\newcommand{\be}{\begin{equation}}
\newcommand{\ee}{\end{equation}}
\newcommand{\bea}{\begin{eqnarray}}
\newcommand{\eea}{\end{eqnarray}}
\def\c#1{~\cite{#1}}
\def\f#1{Fig.~\ref{#1}}

\def\cot{CO$_2$}
\def\h{H$_2$}
\def\water{H$_2$O}

\def\eqq#1{Eq.~(\ref{#1})}

\def\beq{\begin{equation}}
\def\eeq{\end{equation}}

\def\kt{k_{\rm B}T}

\begin{document}

\title{Selective gas capture via kinetic trapping}
\author{Joyjit Kundu}
\email{jkundu@lbl.gov}
\author{Tod Pascal}
\author{David Prendergast}
\author{Stephen Whitelam}
\email{swhitelam@lbl.gov}
\affiliation{Molecular Foundry, Lawrence Berkeley National Laboratory, Berkeley, CA 94720, USA}

%\date{\today}

\begin{abstract}
Conventional approaches to the capture of \cot~by metal-organic frameworks focus on equilibrium conditions, and frameworks that contain little \cot~in equilibrium are often rejected as carbon-capture materials. Here we use a statistical mechanical model, parameterized by quantum mechanical data, to suggest that metal-organic frameworks can be used to separate \cot~from a typical flue gas mixture when used under {\em nonequilibrium} conditions. The origin of this selectivity is an emergent gas-separation mechanism that results from the acquisition by different gas types of different mobilities within a crowded framework. The resulting distribution of gas types within the framework is in general spatially and dynamically heterogeneous. Our results suggest that relaxing the requirement of equilibrium can substantially increase the parameter space of conditions and materials for which selective gas capture can be effected.
\end{abstract}
%\pacs{64.60.De, 64.60.Cn, 05.50.+q}
\maketitle

{\em Introduction.} The burning of carbon-based fossil fuels and the consequent release of CO$_2$ into the atmosphere causes climate change\c{pachauri2014climate}. One technology designed to remove CO$_2$ from the flue (exhaust) gases of power plants is based upon metal-organic frameworks (MOFs), modular crystalline materials whose internal binding sites can host gas molecules\c{queen2014,yaghi2013,yaghi1999,matzger2008,wu2009,poloni_nat,yaghi2013,yaghi2009,long_chemrev,zhou2009,zhou2011}. Standard approaches to understanding gas capture in MOFs focus on equilibrium conditions\c{wu2009,yaghi2005,wu2008,park2012,long2015,david2015,liu2012,Neaton2015}, where the prescription for {\em selective} gas capture is both simple and restrictive: in equilibrium, a framework will harbor \cot~in preference to the other gas types in a mixture if the framework binds more strongly to \cot~than to all the other gas types. For many frameworks, and for flue gas mixtures, this is not the case\c{Neaton2015}. For instance, Mg-MOF-74 is a framework commonly used in the laboratory for gas capture\c{park2012,matzger2008,yaghi2009,krishna2011,valezano2010,bao2011,long_chemrev}. When exposed to \cot~mixed with \water, which is abundant in flue gases, Mg-MOF-74 will, under equilibrium conditions, contain mostly \water\c{piero2013,piero_01,long2015,tan2015competitive}. One response to this problem is to design a material, such as diamine-appended MOF-74\c{david2015,long2015}, better able to capture \cot~in equilibrium. Another response, explored in this paper, is to consider the possibility of doing gas capture under nonequilibrium conditions. 

Gas capture is a dynamic phenomenon\c{piero2013,tan2015competitive,long2012}. Exposed to a MOF, a collection of gas molecules will execute various microscopic processes, including moving through the open space of the framework, and binding to and unbinding from it. In the long-time limit the fraction of a certain gas type resident within the framework is determined by the set of molecule-framework binding affinities, but at intermediate times the composition of gas types within the framework depends in addition on the kinetic parameters that govern the rates of molecular processes\c{bao2011,tan2015competitive,krishna_jpcc,jpcb2005,jiang2008}. Quantum mechanical calculations\c{piero2013} suggest that some frameworks not useful for selective gas capture under equilibrium conditions might perform the same task well under nonequilibrium conditions. For instance, the binding enthalpies of the flue-gas constituents \h, \cot, and \water~in Mg-MOF-74 at $T= 298$ K are $-0.16$~eV, $-0.49$~eV, and $-0.75$~eV, respectively\c{piero2013,valezano2010}. Thus, when exposed to a typical flue-gas mixture (12-15\%~\cot~and 5-7\%~\water\c{compos}), we would expect in equilibrium that most of Mg-MOF-74's binding sites will harbor water molecules (recall that ${\rm eV} \approx 39 \, \kt$ at 298 K). This expectation is consistent with experiment\c{long2015,tan2015competitive}. However, quantum mechanical calculations also indicate that different gas types do not diffuse equally rapidly within the framework. MOF-74 is a three-dimensional structure within which run quasi-one-dimensional channels. Guest molecules binding to open metal sites `coat' the interiors of these channels and restrict the flow of gases through them. In order to move one unit cell down a channel `coated' with molecules of its own kind, a molecule of \h, \cot, or \water~feels an energy barrier of 0.005~eV, 0.04~eV, or 0.06~eV, respectively\c{piero2013}. Thus one might expect \h~to invade the framework first, followed by \cot, followed by \water, with each gas eventually displacing the previous one because of its larger binding affinity for the framework. In this scenario, \cot~could in principle reside within the framework for some time period in a quantity that exceeds its (negligible) equilibrium abundance.

Here we use a statistical mechanical model of gas diffusion and binding within a model framework to confirm this expectation: for a set of three gas types whose hierarchy of (emergent) mobilities is the {\em reverse} of their hierarchy of binding affinities, as for \h, \cot, and \water~in Mg-MOF-74 (this hierarchy of affinities and mobilities is widely observed in many porous materials, e.g. zeolites\c{gang2008} and other MOFs\c{krishna2011,Neaton2015,matzger2011}), a gas that is essentially absent from the framework under equilibrium conditions can be captured under nonequilibrium conditions. The origin of this selective capture is an emergent nonequilibrium `filtration' mechanism that allows, within a crowded framework, certain gas types to invade more rapidly than others. We describe this gas separation mechanism and show that the residence time and the abundance of the desired gas can be increased by {\em impeding} the flow of all gases within the framework, consistent with experiments in which constriction of pore apertures in metal-organic frameworks improved the selectivity of a framework for particular gas types\c{patric2013,yeon2011,gong2009,king2015,zhou2009}.
\begin{figure}
\includegraphics[width=\linewidth]{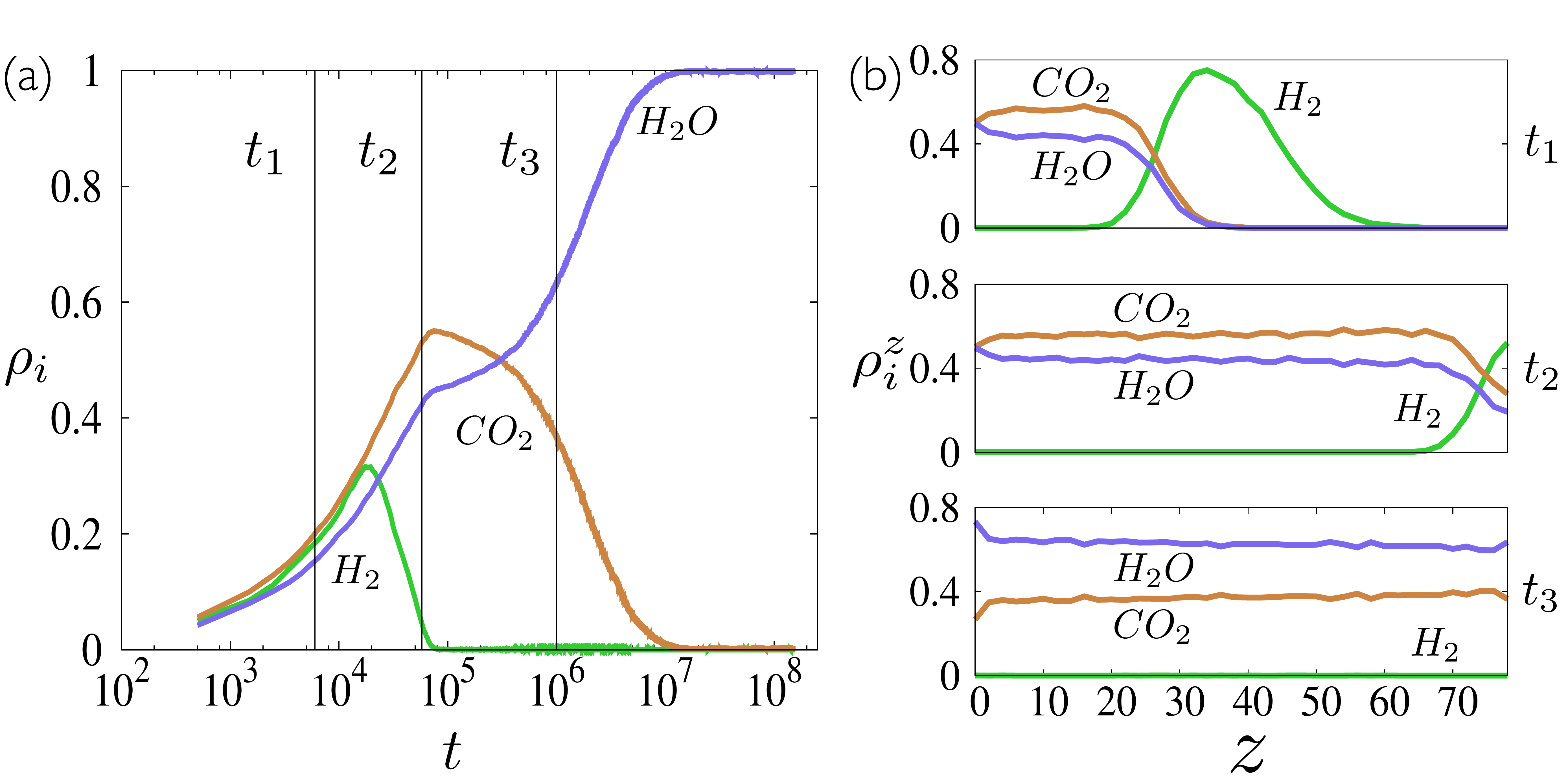}
\caption{Gases absent from a framework in equilibrium can be captured under nonequilibrium conditions. (a) Time evolution of the fraction of binding sites occupied by each gas type in our lattice model of a single Mg-MOF-74 crystal. (b) Fraction of binding sites in the framework occupied by each species as a function of the distance $z$ from the gas reservoir, at three different times. Data were obtained for a lattice of size $L_y=40$, $L_z=80$, at 463 K, averaged over 100 independent simulations. Distances and times are reported in units of $\Delta \ell=1.5 $ \AA~and $\Delta t =10^{-10} $ s, respectively.}
\label{fig1}
\end{figure}

Because of the model's simplicity we do not expect it to be quantitatively precise, but where comparison can be made our results agree qualitatively with experiments\c{long2015,tan2015competitive,jian2012,matzger2011,patric2013}, and indicate that \cot~can under nonequilibrium conditions occupy a substantial fraction of the framework's binding sites. We anticipate that doing gas capture under nonequilibrium conditions will substantially increase the space of protocols and materials for which selective gas capture can be effected.
\begin{figure*}
\includegraphics[width=\linewidth]{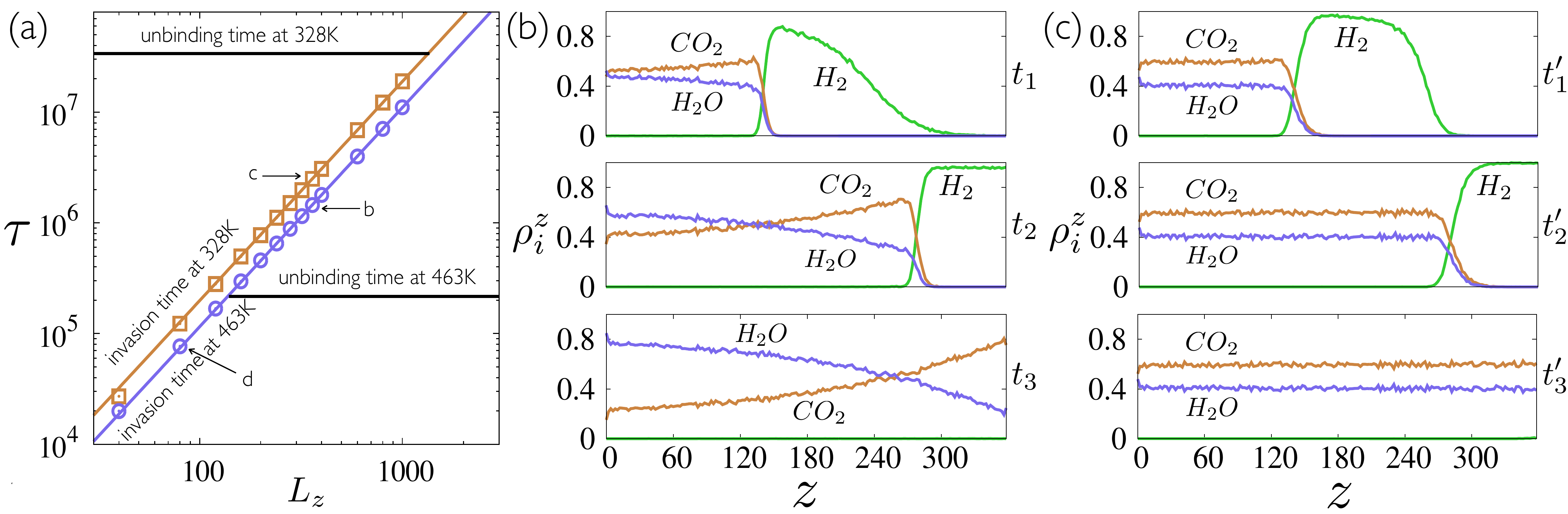}
\caption{(a) The invasion time of \cot~scales with system size, but its unbinding time does not, and thus the nature of the invasion mechanism depends on system size. Simulation data (symbols) can be fit by the sloping lines, respectively $21.71 L_z ^{1.98}$ and $12.62 L_z^{1.98}$ for 328 K and 463 K. Panels (b) and (c) show time-ordered density profiles at state points b and c on panel (a), where the invasion time of \cot~is respectively greater than and less then its unbinding time ($L_y=40$; data are averaged over 100 independent simulations). Point d on panel (a) corresponds to \f{fig1}(b).}
\label{fig2}
\end{figure*}

{\em Model.} We consider a square lattice of $L_y \times L_z$ sites, with periodic boundary conditions imposed in the $y$-(vertical) direction (see \f{fig:fig_SI01}). The column $z=0$ is held in contact with an equimolar~\footnote{In \f{fig:fig_SI02} we show that our qualitative conclusions are unchanged if we take the relative abundance of gas types to be typical of a flue-gas mixture.} reservoir of \h, CO$_2$, and H$_2$O molecules, each represented by a distinct type of particle the size of one lattice site. The boundary $z=L_z-1$ is closed. A site ($i, j$) is called a binding site (denoted by bold circles in \f{fig:fig_SI01}) if $i$ and $j$ are both even. The remaining sites are called free sites, intended to represent empty space. A site can be empty or occupied by a single particle of any type. Particles are hard, and cannot overlap. Particles at free sites do not interact with the framework; particles at binding sites possess a favorable interaction energy of $-0.16$~eV, $-0.49$~eV, or $-0.75$~eV if the particle represents \h, \cot~or \water, respectively\c{piero2013,valezano2010,wu2008}. Particles experience intra-species pairwise nearest-neighbor repulsive interactions of strength 0.0025~eV, 0.02~eV, or 0.03~eV for \h, \cot~or \water, respectively. We impose these interactions so that particles moving along the $z$-axis, past occupied binding sites, experience the energy barriers that particles in Mg-MOF-74 experience as they diffuse along the $c$-axis, though channels `coated' by molecules of the same type\c{piero2013}. We set inter-species nearest-neighbor interactions to be the arithmetic mean of the appropriate intra-species interactions, but we observe little change in our results upon setting inter-species interactions to zero (\f{fig:fig_SI02}). Such robustness indicates that motion through its own species is the process that controls the invasion time of a particular gas type. We assume that all interaction energies are independent of temperature. The basic unit of length is set by the distance between two metal sites ($\approx 3\,$\AA) along the $c$-axis of Mg-MOF-74, corresponding to two lattice units in our model. We therefore set the lattice constant $\Delta \ell=1.5\,$\AA. As in Mg-MOF-74, H$_2$O binds to the framework most strongly but experiences the largest energy barriers to motion along occupied channels, and \h~binds most weakly but experiences the smallest energy barrier to motion along occupied channels. 

We evolved the system using a semi-grand canonical Monte Carlo algorithm that allowed single-particle insertion, removal, and diffusion processes (see SI). The basic timescale $\Delta t$ of our model is set by the time taken for a gas molecule to travel the distance between two sites of the framework. By setting $\Delta t = 10^{-10}$ s we obtain qualitative agreement with experiments that measure the equilibration time of water in a framework pre-loaded with \cot~\cite{tan2015competitive} (see \f{fig:fig_SI03}). Simulations were begun from an empty lattice. We define the density $\rho_j(t)$ of a gas of type $j \in \left\{ \right.$\h,\cot,\water$\left.\right\}$ as the fraction of binding sites occupied at time $t$ by gas type $j$, averaged over many independent simulations. Particles not on binding sites do not contribute to these densities. The resulting dynamics allows particles to enter the framework from its open edge, and to diffuse within and interact with the framework. The lattice is simpler in geometrical terms than MOF-74, which is three dimensional, but it captures the quasi-one-dimensional aspect of gas diffusion within the real structure: we found that dynamics was largely insensitive to the vertical extent $L_y$ of the lattice, because motion of particles into the bulk of the framework is controlled by one-dimensional motion along `channels' adjacent to binding sites (\f{fig:fig_SI04}). In what follows we present distances and times in units of $\Delta \ell=1.5$ \AA~and $\Delta t =10^{-10} $ s, respectively. 

{\em Results.} In \f{fig1}(a) we show the time evolution of the densities of bound gas species within the framework at 463 K. All gas types are bound within the framework for some time period. The bound fractions of \h~and \cot~reach maxima and decline to zero, leaving water to occupy the framework at long times. Thus in equilibrium the bound fraction of \cot~within the framework is effectively zero, as it is in Mg-MOF-74\c{long2015,tan2015competitive}. However, at early times as many as half of the framework's binding sites host a \cot~molecule, i.e. \cot~can be `captured' under nonequilibrium conditions.

\f{fig1}(b) shows at three fixed times the fraction of bound gas as a function of distance $z$ from the gas reservoir. This plot reveals the nature of the nonequilibrium gas-separation mechanism that operates within the framework. It is initially empty. At early times the columns near the reservoir become occupied by an approximately random mixture of all gas types. New gas molecules must pass through these columns in order to enter the framework. As they do so, \water~and \cot~molecules feel a greater energy barrier to their passage than do \h~molecules, and so the latter invade the framework fastest (upper panel). This `filtration' effect is an emergent consequence of the energy barriers felt by particles passing through a crowded framework, which we have parameterized using quantum mechanical data\c{piero2013}. Absent these barriers the gas types diffuse equally rapidly along occupied channels. The gas composition of the framework at early times is spatially heterogeneous. Eventually \cot~and \water~invade the framework and displace the more weakly-binding \h~(middle panel). At long times \cot~is displaced homogeneously by \water~(middle and bottom panels). Eventually, \water~occupies most of the binding sites and the framework equilibrates. 

The spatial distribution of gas types produced by this filtration mechanism depends on system size. The time for a bound molecule to unbind is governed by the ratio of binding enthalpy and temperature, but the invasion time of a gas (e.g. the time taken to occupy 15\% of the sites at the closed end of the lattice) depends both on energetic parameters and on $L_z$. As shown in \f{fig2}(a), the invasion time $\tau$ (in units of $\Delta t$) of \cot~scales as $L_z^{1.98}$ at 328 K and 463 K. Also shown on the plot are the characteristic unbinding times of \cot~at those two temperatures. We see two distinct regimes. In one, the invasion time of \cot~is larger than its unbinding time. Here we find \cot~to be displaced by water in a spatially heterogeneous way, as shown in panel (b), because the open side of the framework approaches equilibrium while \cot~is still invading. In the other regime, illustrated in panel (c), the invasion time of \cot~is smaller than its unbinding time. Here \cot~will reach the closed end of the framework without substantial \cot~unbinding occurring, and it will subsequently be displaced by water in a spatially homogeneous way. 
\begin{figure}
\includegraphics[width=\linewidth]{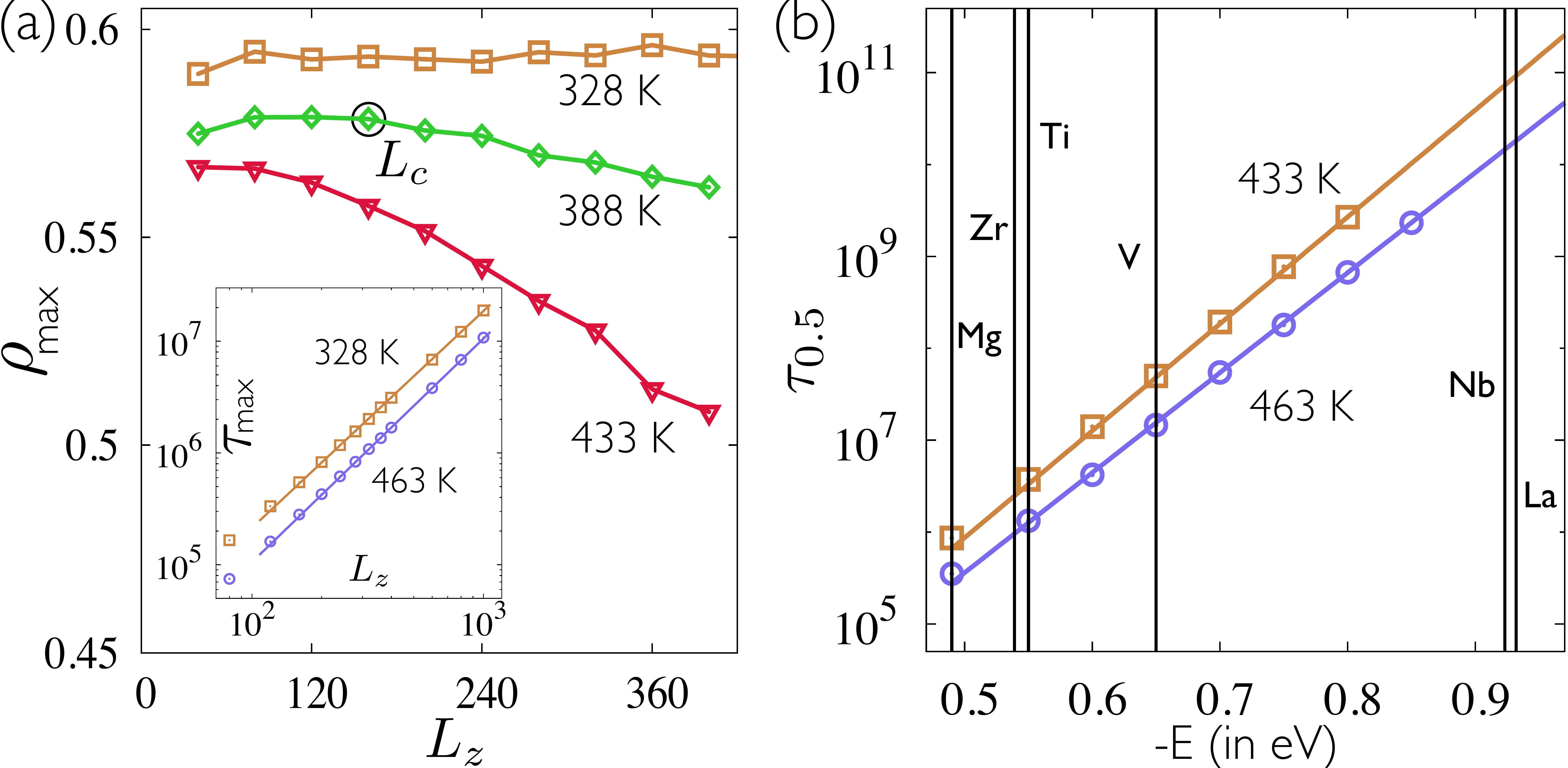}
\caption{(a) Maximum occupancy of \cot~as a function of $L_z$ at different temperatures. The label $L_{\rm c}$ indicates the `crossover lengthscale' (at 388 K) at which invasion and unbinding times of \cot~are comparable. Inset: time of maximum occupancy of \cot~as a function of the linear extent $L_z$ of the framework at two different temperatures. The straight lines through the data for 328 K and 463 K have equations $t_{\rm max}=28.18~ L_z^{1.94}$ and $t_{\rm max}=10.55~ L_z^{2.00}$, respectively. Here $L_y=40$. (b) Time at which half of all binding sites harbor \cot, as a function of the binding enthalpy of \cot, at 433 K and 463 K (here $L_y=10$ and $L_z=100$). Straight lines are Arrhenius fits: $\tau_{0.5}=\mbox{A} \exp(-E/{k_B}T)$, where $\mbox{A(433 K)}\approx 1.31$ and $\mbox{A(463 K)}\approx 1.30$. The vertical lines correspond to different metals (in MOF-74) with different binding enthalpies for \cot~(all have higher binding affinity for \water)\c{piero_01,poloni_jpcl}. Here we fix the binding enthalpies of \h~and \water~to be $-0.15$ eV and $-2.00$ eV respectively.}
\label{fig3}
\end{figure}

The maximum abundance $\rho_{\rm max}$ of bound \cot~within the framework (at any time) also depends on $L_z$: it is insensitive to $L_z$ for $L_z$ small enough that \cot's invasion time is smaller than its unbinding time, and diminishes with $L_z$ for $L_z$ large enough that \cot's invasion time is larger than its unbinding time: see \f{fig3}(a). The `crossover' length $L_{\rm c}$ separating these two regimes depends strongly upon temperature, scaling roughly as $L_{\rm c} \sim \exp(\beta E/q)$, where $E$ is the binding enthalpy of \cot, and $q \approx 2$ (varying weakly with $T$). Gas capture experiments often use powders whose grain sizes are broadly distributed\c{bao2011}. A simple strategy to maximize gas uptake by a powder is to ensure that grains' characteristic lengths are smaller than $L_{\rm c}$. In \f{fig:fig_SI05} and \f{fig:fig_SI06} we show that \cot~capture can be effected in a range of grain sizes simultaneously. Inset of \f{fig3}(a) shows that the time at which the maximum abundance of \cot~is attained increases as a power of $L_z$. At 298 K, a framework with $L_z=L_{\rm c}\approx 0.43\,\mu \mbox{m}$ will harbor \cot~at more than 40\% of its binding sites up to a time of $\sim 0.1$ s (see \f{fig:fig_SI06}). 

The timescale on which nonequilibrium capture can be achieved can be increased through choice of metal constituents of MOF-74 that bind \cot~more strongly than does Mg. This is true even if those metals bind water more strongly; thus, the requirements for gas capture out of equilibrium are less restrictive than for capture in equilibrium: see~\f{fig3}(b). For instance, a single MOF-74 crystal of length $0.43~\mu \mbox{m}$, made from a metal whose binding enthalpy with \cot~is $-0.70$ eV, which is experimentally realizable, can harbor \cot~up to a time of $\sim460$ s (at 298 K). At that temperature the crossover length, the grain size limit at which nonequilibrium uptake capacity is maximum, is $L_{\rm c} \approx 26 \, \mu \mbox{m}$.

We note finally that the time `window' of \cot~capture can also be enlarged by increasing the separation of timescales associated with the nonequilibrium `filtration' mechanism, and that this in turn can be achieved by {\em impeding} the flow of gases into the framework: see \f{fig:fig_SI07}. Constriction of pore apertures in metal-organic frameworks has been seen to improve the selectivity of a framework for particular gas types in experiment\c{patric2013,yeon2011,gong2009,king2015,zhou2009} and atomistic simulation\c{liu2008,liu2009}. 

{\em Conclusions.}  We have studied a simple model of gas-framework dynamics inspired by experiment and parameterized using quantum mechanical data. For a set of three gas types whose hierarchy of mobilities in a crowded environment is the reverse of their hierarchy of framework-binding affinities, a gas (\cot) that is essentially absent from the framework under equilibrium conditions can be captured under nonequilibrium conditions. To make precise predictions for specific experiments it may be necessary to relax several of the simplifying assumptions that we have made. For instance, we have neglected attractive molecule-molecule interactions, which at sufficiently low temperature may induce condensation or phase coexistence within the framework\c{barun2015}. We have also considered the existence of only one kind of binding site, although in Mg-MOF-74 the displacement of \cot~by \water~may involve the passing of \cot~from the primary binding site to a secondary one, through a low-energy exchange pathway\c{tan2015competitive,piero2013,piero_02,queen2011} (note though that \cot~binds almost as strongly to the secondary site as the primary one in Mg-MOF-74\c{piero_02}, indicating that energy barriers for its removal from the framework are similar to those assumed here). Nonetheless, our results agree qualitatively with existing experimental observations: water inhabits Mg-MOF-74 in preference to \cot~in equilibrium\c{long2015,tan2015competitive}; \cot~can be resident within the Mg-MOF-74 for some considerable time away from equilibrium\c{tan2015competitive}; and narrower pores lead to better gas-capture selectivity\c{patric2013,yeon2011,gong2009,king2015}. The nonequilibrium `filtration' mechanism seen in \f{fig1} and \f{fig2} provides a possible microscopic explanation for this latter phenomenon.
 
{\em Acknowledgements.} We thank Pieremanuele Canepa and Rebecca Siegelman for discussions, and Jeff Martell for comments on the manuscript. JK was supported by the Center for Gas Separations Relevant to Clean Energy Technologies, an Energy Frontier Research Center funded by the U.S. Department of Energy, Office of Science, Basic Energy Sciences under award DE-SC0001015. DGP and SW were partially supported by the same Center. Work at the Molecular Foundry was supported by the Office of Science, Office of Basic Energy Sciences, of the U.S. Department of Energy under Contract No. DE-AC02-05CH11231.

%\bibliography{ref01}

\begin{thebibliography}{41}%
\makeatletter
\providecommand \@ifxundefined [1]{%
 \@ifx{#1\undefined}
}%
\providecommand \@ifnum [1]{%
 \ifnum #1\expandafter \@firstoftwo
 \else \expandafter \@secondoftwo
 \fi
}%
\providecommand \@ifx [1]{%
 \ifx #1\expandafter \@firstoftwo
 \else \expandafter \@secondoftwo
 \fi
}%
\providecommand \natexlab [1]{#1}%
\providecommand \enquote  [1]{``#1''}%
\providecommand \bibnamefont  [1]{#1}%
\providecommand \bibfnamefont [1]{#1}%
\providecommand \citenamefont [1]{#1}%
\providecommand \href@noop [0]{\@secondoftwo}%
\providecommand \href [0]{\begingroup \@sanitize@url \@href}%
\providecommand \@href[1]{\@@startlink{#1}\@@href}%
\providecommand \@@href[1]{\endgroup#1\@@endlink}%
\providecommand \@sanitize@url [0]{\catcode `\\12\catcode `\$12\catcode
  `\&12\catcode `\#12\catcode `\^12\catcode `\_12\catcode `\%12\relax}%
\providecommand \@@startlink[1]{}%
\providecommand \@@endlink[0]{}%
\providecommand \url  [0]{\begingroup\@sanitize@url \@url }%
\providecommand \@url [1]{\endgroup\@href {#1}{\urlprefix }}%
\providecommand \urlprefix  [0]{URL }%
\providecommand \Eprint [0]{\href }%
\providecommand \doibase [0]{http://dx.doi.org/}%
\providecommand \selectlanguage [0]{\@gobble}%
\providecommand \bibinfo  [0]{\@secondoftwo}%
\providecommand \bibfield  [0]{\@secondoftwo}%
\providecommand \translation [1]{[#1]}%
\providecommand \BibitemOpen [0]{}%
\providecommand \bibitemStop [0]{}%
\providecommand \bibitemNoStop [0]{.\EOS\space}%
\providecommand \EOS [0]{\spacefactor3000\relax}%
\providecommand \BibitemShut  [1]{\csname bibitem#1\endcsname}%
\let\auto@bib@innerbib\@empty
%</preamble>
\bibitem [{\citenamefont {Pachauri}\ \emph {et~al.}(2014)\citenamefont
  {Pachauri}, \citenamefont {Allen}, \citenamefont {Barros}, \citenamefont
  {Broome}, \citenamefont {Cramer}, \citenamefont {Christ}, \citenamefont
  {Church}, \citenamefont {Clarke}, \citenamefont {Dahe}, \citenamefont
  {Dasgupta} \emph {et~al.}}]{pachauri2014climate}%
  \BibitemOpen
  \bibfield  {author} {\bibinfo {author} {\bibfnamefont {R.~K.}\ \bibnamefont
  {Pachauri}}, \bibinfo {author} {\bibfnamefont {M.~R.}~\bibnamefont {Allen}},
  \bibinfo {author} {\bibfnamefont {V.~R.}~\bibnamefont {Barros}}, \bibinfo
  {author} {\bibfnamefont {J.}~\bibnamefont {Broome}}, \bibinfo {author}
  {\bibfnamefont {W.}~\bibnamefont {Cramer}}, \bibinfo {author} {\bibfnamefont
  {R.}~\bibnamefont {Christ}}, \bibinfo {author} {\bibfnamefont
  {J.~A.}~\bibnamefont {Church}}, \bibinfo {author} {\bibfnamefont
  {L.}~\bibnamefont {Clarke}}, \bibinfo {author} {\bibfnamefont
  {Q.}~\bibnamefont {Dahe}}, \bibinfo {author} {\bibfnamefont {P.}~\bibnamefont
  {Dasgupta}},  \emph {et~al.},\ }\href@noop {} {\  (\bibinfo {year}
  {2014})}\BibitemShut {NoStop}%
\bibitem [{\citenamefont {Queen}\ \emph {et~al.}(2014)\citenamefont {Queen},
  \citenamefont {Hudson}, \citenamefont {Bloch}, \citenamefont {Mason},
  \citenamefont {Gonzalez}, \citenamefont {Lee}, \citenamefont {Gygi},
  \citenamefont {Howe}, \citenamefont {Lee}, \citenamefont {Darwish},
  \citenamefont {James}, \citenamefont {Peterson}, \citenamefont {Teat},
  \citenamefont {Smit}, \citenamefont {Neaton}, \citenamefont {Long},\ and\
  \citenamefont {Brown}}]{queen2014}%
  \BibitemOpen
  \bibfield  {author} {\bibinfo {author} {\bibfnamefont {W.~L.}\ \bibnamefont
  {Queen}}, \bibinfo {author} {\bibfnamefont {M.~R.}\ \bibnamefont {Hudson}},
  \bibinfo {author} {\bibfnamefont {E.~D.}\ \bibnamefont {Bloch}}, \bibinfo
  {author} {\bibfnamefont {J.~A.}\ \bibnamefont {Mason}}, \bibinfo {author}
  {\bibfnamefont {M.~I.}\ \bibnamefont {Gonzalez}}, \bibinfo {author}
  {\bibfnamefont {J.~S.}\ \bibnamefont {Lee}}, \bibinfo {author} {\bibfnamefont
  {D.}~\bibnamefont {Gygi}}, \bibinfo {author} {\bibfnamefont {J.~D.}\
  \bibnamefont {Howe}}, \bibinfo {author} {\bibfnamefont {K.}~\bibnamefont
  {Lee}}, \bibinfo {author} {\bibfnamefont {T.~A.}\ \bibnamefont {Darwish}},
  \bibinfo {author} {\bibfnamefont {M.}~\bibnamefont {James}}, \bibinfo
  {author} {\bibfnamefont {V.~K.}\ \bibnamefont {Peterson}}, \bibinfo {author}
  {\bibfnamefont {S.~J.}\ \bibnamefont {Teat}}, \bibinfo {author}
  {\bibfnamefont {B.}~\bibnamefont {Smit}}, \bibinfo {author} {\bibfnamefont
  {J.~B.}\ \bibnamefont {Neaton}}, \bibinfo {author} {\bibfnamefont {J.~R.}\
  \bibnamefont {Long}}, \ and\ \bibinfo {author} {\bibfnamefont {C.~M.}\
  \bibnamefont {Brown}},\ }\href@noop {} {\bibfield  {journal} {\bibinfo
  {journal} {Chem. Sci.}\ }\textbf {\bibinfo {volume} {5}},\ \bibinfo {pages}
  {4569} (\bibinfo {year} {2014})}\BibitemShut {NoStop}%
\bibitem [{\citenamefont {Furukawa}\ \emph {et~al.}(2013)\citenamefont
  {Furukawa}, \citenamefont {Cordova}, \citenamefont {O'Keeffe},\ and\
  \citenamefont {Yaghi}}]{yaghi2013}%
  \BibitemOpen
  \bibfield  {author} {\bibinfo {author} {\bibfnamefont {H.}~\bibnamefont
  {Furukawa}}, \bibinfo {author} {\bibfnamefont {K.~E.}\ \bibnamefont
  {Cordova}}, \bibinfo {author} {\bibfnamefont {M.}~\bibnamefont {O'Keeffe}}, \
  and\ \bibinfo {author} {\bibfnamefont {O.~M.}\ \bibnamefont {Yaghi}},\
  }\href@noop {} {\bibfield  {journal} {\bibinfo  {journal} {Science}\ }\textbf
  {\bibinfo {volume} {341}},\ \bibinfo {pages} {1230444} (\bibinfo {year}
  {2013})}\BibitemShut {NoStop}%
\bibitem [{\citenamefont {Li}\ \emph {et~al.}(1999)\citenamefont {Li},
  \citenamefont {Eddaoudi}, \citenamefont {O'Keeffe},\ and\ \citenamefont
  {Yaghi}}]{yaghi1999}%
  \BibitemOpen
  \bibfield  {author} {\bibinfo {author} {\bibfnamefont {H.}~\bibnamefont
  {Li}}, \bibinfo {author} {\bibfnamefont {M.}~\bibnamefont {Eddaoudi}},
  \bibinfo {author} {\bibfnamefont {M.}~\bibnamefont {O'Keeffe}}, \ and\
  \bibinfo {author} {\bibfnamefont {O.~M.}\ \bibnamefont {Yaghi}},\ }\href@noop
  {} {\bibfield  {journal} {\bibinfo  {journal} {Nature}\ }\textbf {\bibinfo
  {volume} {402}},\ \bibinfo {pages} {276} (\bibinfo {year}
  {1999})}\BibitemShut {NoStop}%
\bibitem [{\citenamefont {Caskey}\ \emph {et~al.}(2008)\citenamefont {Caskey},
  \citenamefont {Wong-Foy},\ and\ \citenamefont {Matzger}}]{matzger2008}%
  \BibitemOpen
  \bibfield  {author} {\bibinfo {author} {\bibfnamefont {S.~R.}\ \bibnamefont
  {Caskey}}, \bibinfo {author} {\bibfnamefont {A.~G.}\ \bibnamefont
  {Wong-Foy}}, \ and\ \bibinfo {author} {\bibfnamefont {A.~J.}\ \bibnamefont
  {Matzger}},\ }\href@noop {} {\bibfield  {journal} {\bibinfo  {journal} {J.
  Am. Chem. Soc.}\ }\textbf {\bibinfo {volume} {130}},\ \bibinfo {pages}
  {10870} (\bibinfo {year} {2008})}\BibitemShut {NoStop}%
\bibitem [{\citenamefont {Wu}\ \emph {et~al.}(2009)\citenamefont {Wu},
  \citenamefont {Zhou},\ and\ \citenamefont {Yildirim}}]{wu2009}%
  \BibitemOpen
  \bibfield  {author} {\bibinfo {author} {\bibfnamefont {H.}~\bibnamefont
  {Wu}}, \bibinfo {author} {\bibfnamefont {W.}~\bibnamefont {Zhou}}, \ and\
  \bibinfo {author} {\bibfnamefont {T.}~\bibnamefont {Yildirim}},\ }\href@noop
  {} {\bibfield  {journal} {\bibinfo  {journal} {J. Am. Chem. Soc.}\ }\textbf
  {\bibinfo {volume} {131}},\ \bibinfo {pages} {4995} (\bibinfo {year}
  {2009})}\BibitemShut {NoStop}%
\bibitem [{\citenamefont {Dzubak}\ \emph {et~al.}(2012)\citenamefont {Dzubak},
  \citenamefont {Lin}, \citenamefont {Kim}, \citenamefont {Swisher},
  \citenamefont {Poloni}, \citenamefont {Maximoff}, \citenamefont {Smit},\ and\
  \citenamefont {Gagliardi}}]{poloni_nat}%
  \BibitemOpen
  \bibfield  {author} {\bibinfo {author} {\bibfnamefont {A.~L.}\ \bibnamefont
  {Dzubak}}, \bibinfo {author} {\bibfnamefont {L.-C.}\ \bibnamefont {Lin}},
  \bibinfo {author} {\bibfnamefont {J.}~\bibnamefont {Kim}}, \bibinfo {author}
  {\bibfnamefont {J.~A.}\ \bibnamefont {Swisher}}, \bibinfo {author}
  {\bibfnamefont {R.}~\bibnamefont {Poloni}}, \bibinfo {author} {\bibfnamefont
  {S.~N.}\ \bibnamefont {Maximoff}}, \bibinfo {author} {\bibfnamefont
  {B.}~\bibnamefont {Smit}}, \ and\ \bibinfo {author} {\bibfnamefont
  {L.}~\bibnamefont {Gagliardi}},\ }\href@noop {} {\bibfield  {journal}
  {\bibinfo  {journal} {Nat. Chem.}\ }\textbf {\bibinfo {volume} {4}},\
  \bibinfo {pages} {810} (\bibinfo {year} {2012})}\BibitemShut {NoStop}%
\bibitem [{\citenamefont {Britt}\ \emph {et~al.}(2009)\citenamefont {Britt},
  \citenamefont {Furukawa}, \citenamefont {Wang}, \citenamefont {Glover},\ and\
  \citenamefont {Yaghi}}]{yaghi2009}%
  \BibitemOpen
  \bibfield  {author} {\bibinfo {author} {\bibfnamefont {D.}~\bibnamefont
  {Britt}}, \bibinfo {author} {\bibfnamefont {H.}~\bibnamefont {Furukawa}},
  \bibinfo {author} {\bibfnamefont {B.}~\bibnamefont {Wang}}, \bibinfo {author}
  {\bibfnamefont {T.~G.}\ \bibnamefont {Glover}}, \ and\ \bibinfo {author}
  {\bibfnamefont {O.~M.}\ \bibnamefont {Yaghi}},\ }\href@noop {} {\bibfield
  {journal} {\bibinfo  {journal} {Proc. Natl. Acad. Sci.}\ }\textbf {\bibinfo
  {volume} {106}},\ \bibinfo {pages} {20637} (\bibinfo {year}
  {2009})}\BibitemShut {NoStop}%
\bibitem [{\citenamefont {Sumida}\ \emph {et~al.}(2012)\citenamefont {Sumida},
  \citenamefont {Rogow}, \citenamefont {Mason}, \citenamefont {McDonald},
  \citenamefont {Bloch}, \citenamefont {Herm}, \citenamefont {Bae},\ and\
  \citenamefont {Long}}]{long_chemrev}%
  \BibitemOpen
  \bibfield  {author} {\bibinfo {author} {\bibfnamefont {K.}~\bibnamefont
  {Sumida}}, \bibinfo {author} {\bibfnamefont {D.~L.}\ \bibnamefont {Rogow}},
  \bibinfo {author} {\bibfnamefont {J.~A.}\ \bibnamefont {Mason}}, \bibinfo
  {author} {\bibfnamefont {T.~M.}\ \bibnamefont {McDonald}}, \bibinfo {author}
  {\bibfnamefont {E.~D.}\ \bibnamefont {Bloch}}, \bibinfo {author}
  {\bibfnamefont {Z.~R.}\ \bibnamefont {Herm}}, \bibinfo {author}
  {\bibfnamefont {T.-H.}\ \bibnamefont {Bae}}, \ and\ \bibinfo {author}
  {\bibfnamefont {J.~R.}\ \bibnamefont {Long}},\ }\href@noop {} {\bibfield
  {journal} {\bibinfo  {journal} {Chem. Rev.}\ }\textbf {\bibinfo {volume}
  {112}},\ \bibinfo {pages} {724} (\bibinfo {year} {2012})}\BibitemShut
  {NoStop}%
\bibitem [{\citenamefont {Li}\ \emph {et~al.}(2009{\natexlab{a}})\citenamefont
  {Li}, \citenamefont {Kuppler},\ and\ \citenamefont {Zhou}}]{zhou2009}%
  \BibitemOpen
  \bibfield  {author} {\bibinfo {author} {\bibfnamefont {J.-R.}\ \bibnamefont
  {Li}}, \bibinfo {author} {\bibfnamefont {R.~J.}\ \bibnamefont {Kuppler}}, \
  and\ \bibinfo {author} {\bibfnamefont {H.-C.}\ \bibnamefont {Zhou}},\
  }\href@noop {} {\bibfield  {journal} {\bibinfo  {journal} {Chem. Soc. Rev.}\
  }\textbf {\bibinfo {volume} {38}},\ \bibinfo {pages} {1477} (\bibinfo {year}
  {2009}{\natexlab{a}})}\BibitemShut {NoStop}%
\bibitem [{\citenamefont {Li}\ \emph {et~al.}(2011)\citenamefont {Li},
  \citenamefont {Ma}, \citenamefont {McCarthy}, \citenamefont
  {Sculley}, \citenamefont {Yu}, \citenamefont {Jeong}, \citenamefont {Balbuena},\ and\
  \citenamefont {Zhou}}]{zhou2011}%
  \BibitemOpen
  \bibfield  {author} {\bibinfo {author} {\bibfnamefont {J.-R.}\ \bibnamefont
  {Li}}, \bibinfo {author} {\bibfnamefont {Y.}~\bibnamefont {Ma}}, \bibinfo
  {author} {\bibfnamefont {M.~C.}\ \bibnamefont {McCarthy}}, \bibinfo {author}
  {\bibfnamefont {J.}\ \bibnamefont {Sculley}}, 
  \bibinfo {author} {\bibfnamefont {J.}~\bibnamefont {Yu}}, \bibinfo {author}
  {\bibfnamefont {H.-K.}\ \bibnamefont {Jeong}}, \bibinfo {author}
  {\bibfnamefont {P.~B.}\ \bibnamefont {Balbuena}}, \ and\ \bibinfo {author}
  {\bibfnamefont {H.-C.}\ \bibnamefont {Zhou}},\ }\href@noop {} {\bibfield
  {journal} {\bibinfo  {journal} {Coord. Chem. Rev.}\ }\textbf {\bibinfo
  {volume} {255}},\ \bibinfo {pages} {1791} (\bibinfo {year}
  {2011})}\BibitemShut {NoStop}%
\bibitem [{\citenamefont {Millward}\ and\ \citenamefont
  {Yaghi}(2005)}]{yaghi2005}%
  \BibitemOpen
  \bibfield  {author} {\bibinfo {author} {\bibfnamefont {A.~R.}\ \bibnamefont
  {Millward}}\ and\ \bibinfo {author} {\bibfnamefont {O.~M.}\ \bibnamefont
  {Yaghi}},\ }\href@noop {} {\bibfield  {journal} {\bibinfo  {journal} {J. Am.
  Chem. Soc.}\ }\textbf {\bibinfo {volume} {127}},\ \bibinfo {pages} {17998}
  (\bibinfo {year} {2005})}\BibitemShut {NoStop}%
\bibitem [{\citenamefont {Zhou}\ \emph {et~al.}(2008)\citenamefont {Zhou},
  \citenamefont {Wu},\ and\ \citenamefont {Yildirim}}]{wu2008}%
  \BibitemOpen
  \bibfield  {author} {\bibinfo {author} {\bibfnamefont {W.}~\bibnamefont
  {Zhou}}, \bibinfo {author} {\bibfnamefont {H.}~\bibnamefont {Wu}}, \ and\
  \bibinfo {author} {\bibfnamefont {T.}~\bibnamefont {Yildirim}},\ }\href@noop
  {} {\bibfield  {journal} {\bibinfo  {journal} {J. Am. Chem. Soc.}\ }\textbf
  {\bibinfo {volume} {130}},\ \bibinfo {pages} {15268} (\bibinfo {year}
  {2008})}\BibitemShut {NoStop}%
\bibitem [{\citenamefont {Park}\ \emph {et~al.}(2012)\citenamefont {Park},
  \citenamefont {Kim}, \citenamefont {Han},\ and\ \citenamefont
  {Jung}}]{park2012}%
  \BibitemOpen
  \bibfield  {author} {\bibinfo {author} {\bibfnamefont {J.}~\bibnamefont
  {Park}}, \bibinfo {author} {\bibfnamefont {H.}~\bibnamefont {Kim}}, \bibinfo
  {author} {\bibfnamefont {S.~S.}\ \bibnamefont {Han}}, \ and\ \bibinfo
  {author} {\bibfnamefont {Y.}~\bibnamefont {Jung}},\ }\href@noop {} {\bibfield
   {journal} {\bibinfo  {journal} {J. Phys. Chem. Lett.}\ }\textbf {\bibinfo
  {volume} {3}},\ \bibinfo {pages} {826} (\bibinfo {year} {2012})}\BibitemShut
  {NoStop}%
\bibitem [{\citenamefont {Mason}\ \emph {et~al.}(2015)\citenamefont {Mason},
  \citenamefont {McDonald}, \citenamefont {Bae}, \citenamefont {Bachman},
  \citenamefont {Sumida}, \citenamefont {Dutton}, \citenamefont {Kaye},\ and\
  \citenamefont {Long}}]{long2015}%
  \BibitemOpen
  \bibfield  {author} {\bibinfo {author} {\bibfnamefont {J.~A.}\ \bibnamefont
  {Mason}}, \bibinfo {author} {\bibfnamefont {T.~M.}\ \bibnamefont {McDonald}},
  \bibinfo {author} {\bibfnamefont {T.-H.}\ \bibnamefont {Bae}}, \bibinfo
  {author} {\bibfnamefont {J.~E.}\ \bibnamefont {Bachman}}, \bibinfo {author}
  {\bibfnamefont {K.}~\bibnamefont {Sumida}}, \bibinfo {author} {\bibfnamefont
  {J.~J.}\ \bibnamefont {Dutton}}, \bibinfo {author} {\bibfnamefont {S.~S.}\
  \bibnamefont {Kaye}}, \ and\ \bibinfo {author} {\bibfnamefont {J.~R.}\
  \bibnamefont {Long}},\ }\href@noop {} {\bibfield  {journal} {\bibinfo
  {journal} {J. Am. Chem. Soc.}\ }\textbf {\bibinfo {volume} {137}},\ \bibinfo
  {pages} {4787} (\bibinfo {year} {2015})}\BibitemShut {NoStop}%
\bibitem [{\citenamefont {McDonald}\ \emph {et~al.}(2015)\citenamefont
  {McDonald}, \citenamefont {Mason}, \citenamefont {Kong}, \citenamefont
  {Bloch}, \citenamefont {Gygi}, \citenamefont {Dani}, \citenamefont
  {Crocell\`{a}}, \citenamefont {Giordanino}, \citenamefont {Odoh},
  \citenamefont {Drisdell}, \citenamefont {Vlaisavljevich}, \citenamefont
  {Dzubak}, \citenamefont {Poloni}, \citenamefont {Schnell}, \citenamefont
  {Planas}, \citenamefont {Lee}, \citenamefont {Pascal}, \citenamefont {Wan},
  \citenamefont {Prendergast}, \citenamefont {Neaton}, \citenamefont {Smit},
  \citenamefont {Kortright}, \citenamefont {Gagliardi}, \citenamefont
  {Bordiga}, \citenamefont {Reimer},\ and\ \citenamefont {Long}}]{david2015}%
  \BibitemOpen
  \bibfield  {author} {\bibinfo {author} {\bibfnamefont {T.~M.}\ \bibnamefont
  {McDonald}}, \bibinfo {author} {\bibfnamefont {J.~A.}\ \bibnamefont {Mason}},
  \bibinfo {author} {\bibfnamefont {X.}~\bibnamefont {Kong}}, \bibinfo {author}
  {\bibfnamefont {E.~D.}\ \bibnamefont {Bloch}}, \bibinfo {author}
  {\bibfnamefont {D.}~\bibnamefont {Gygi}}, \bibinfo {author} {\bibfnamefont
  {A.}~\bibnamefont {Dani}}, \bibinfo {author} {\bibfnamefont {V.}~\bibnamefont
  {Crocell\`{a}}}, \bibinfo {author} {\bibfnamefont {F.}~\bibnamefont
  {Giordanino}}, \bibinfo {author} {\bibfnamefont {S.~O.}\ \bibnamefont
  {Odoh}}, \bibinfo {author} {\bibfnamefont {W.~S.}\ \bibnamefont {Drisdell}},
  \bibinfo {author} {\bibfnamefont {B.}~\bibnamefont {Vlaisavljevich}},
  \bibinfo {author} {\bibfnamefont {A.~L.}\ \bibnamefont {Dzubak}}, \bibinfo
  {author} {\bibfnamefont {R.}~\bibnamefont {Poloni}}, \bibinfo {author}
  {\bibfnamefont {S.~K.}\ \bibnamefont {Schnell}}, \bibinfo {author}
  {\bibfnamefont {N.}~\bibnamefont {Planas}}, \bibinfo {author} {\bibfnamefont
  {K.}~\bibnamefont {Lee}}, \bibinfo {author} {\bibfnamefont {T.}~\bibnamefont
  {Pascal}}, \bibinfo {author} {\bibfnamefont {L.~F.}\ \bibnamefont {Wan}},
  \bibinfo {author} {\bibfnamefont {D.}~\bibnamefont {Prendergast}}, \bibinfo
  {author} {\bibfnamefont {J.~B.}\ \bibnamefont {Neaton}}, \bibinfo {author}
  {\bibfnamefont {B.}~\bibnamefont {Smit}}, \bibinfo {author} {\bibfnamefont
  {J.~B.}\ \bibnamefont {Kortright}}, \bibinfo {author} {\bibfnamefont
  {L.}~\bibnamefont {Gagliardi}}, \bibinfo {author} {\bibfnamefont
  {S.}~\bibnamefont {Bordiga}}, \bibinfo {author} {\bibfnamefont {J.~A.}\
  \bibnamefont {Reimer}}, \ and\ \bibinfo {author} {\bibfnamefont {J.~R.}\
  \bibnamefont {Long}},\ }\href@noop {} {\bibfield  {journal} {\bibinfo
  {journal} {Nature}\ }\textbf {\bibinfo {volume} {519}},\ \bibinfo {pages}
  {303} (\bibinfo {year} {2015})}\BibitemShut {NoStop}%
\bibitem [{\citenamefont {Liu}\ \emph {et~al.}(2012{\natexlab{a}})\citenamefont
  {Liu}, \citenamefont {Thallapally}, \citenamefont {McGrail}, \citenamefont
  {Brown},\ and\ \citenamefont {Liu}}]{liu2012}%
  \BibitemOpen
  \bibfield  {author} {\bibinfo {author} {\bibfnamefont {J.}~\bibnamefont
  {Liu}}, \bibinfo {author} {\bibfnamefont {P.~K.}\ \bibnamefont
  {Thallapally}}, \bibinfo {author} {\bibfnamefont {B.~P.}\ \bibnamefont
  {McGrail}}, \bibinfo {author} {\bibfnamefont {D.~R.}\ \bibnamefont {Brown}},
  \ and\ \bibinfo {author} {\bibfnamefont {J.}~\bibnamefont {Liu}},\
  }\href@noop {} {\bibfield  {journal} {\bibinfo  {journal} {Chem. Soc. Rev.}\
  }\textbf {\bibinfo {volume} {41}},\ \bibinfo {pages} {2308} (\bibinfo {year}
  {2012}{\natexlab{a}})}\BibitemShut {NoStop}%
\bibitem [{\citenamefont {Lee}\ \emph {et~al.}(2015)\citenamefont {Lee},
  \citenamefont {Howe}, \citenamefont {Lin}, \citenamefont {Smit},\ and\
  \citenamefont {Neaton}}]{Neaton2015}%
  \BibitemOpen
  \bibfield  {author} {\bibinfo {author} {\bibfnamefont {K.}~\bibnamefont
  {Lee}}, \bibinfo {author} {\bibfnamefont {J.~D.}\ \bibnamefont {Howe}},
  \bibinfo {author} {\bibfnamefont {L.-C.}\ \bibnamefont {Lin}}, \bibinfo
  {author} {\bibfnamefont {B.}~\bibnamefont {Smit}}, \ and\ \bibinfo {author}
  {\bibfnamefont {J.~B.}\ \bibnamefont {Neaton}},\ }\href@noop {} {\bibfield
  {journal} {\bibinfo  {journal} {Chem. Mater.}\ }\textbf {\bibinfo {volume}
  {27}},\ \bibinfo {pages} {668} (\bibinfo {year} {2015})}\BibitemShut
  {NoStop}%
\bibitem [{\citenamefont {Mason}\ \emph {et~al.}(2011)\citenamefont {Mason},
  \citenamefont {Sumida}, \citenamefont {Herm}, \citenamefont {Krishna},\ and\
  \citenamefont {Long}}]{krishna2011}%
  \BibitemOpen
  \bibfield  {author} {\bibinfo {author} {\bibfnamefont {J.~A.}\ \bibnamefont
  {Mason}}, \bibinfo {author} {\bibfnamefont {K.}~\bibnamefont {Sumida}},
  \bibinfo {author} {\bibfnamefont {Z.~R.}\ \bibnamefont {Herm}}, \bibinfo
  {author} {\bibfnamefont {R.}~\bibnamefont {Krishna}}, \ and\ \bibinfo
  {author} {\bibfnamefont {J.~R.}\ \bibnamefont {Long}},\ }\href@noop {}
  {\bibfield  {journal} {\bibinfo  {journal} {Energy Environ. Sci.}\ }\textbf
  {\bibinfo {volume} {4}},\ \bibinfo {pages} {3030} (\bibinfo {year}
  {2011})}\BibitemShut {NoStop}%
\bibitem [{\citenamefont {Valenzano}\ \emph {et~al.}(2010)\citenamefont
  {Valenzano}, \citenamefont {Civalleri}, \citenamefont {Chavan}, \citenamefont
  {Palomino}, \citenamefont {Are\`{a}n},\ and\ \citenamefont
  {Bordiga}}]{valezano2010}%
  \BibitemOpen
  \bibfield  {author} {\bibinfo {author} {\bibfnamefont {L.}~\bibnamefont
  {Valenzano}}, \bibinfo {author} {\bibfnamefont {B.}~\bibnamefont
  {Civalleri}}, \bibinfo {author} {\bibfnamefont {S.}~\bibnamefont {Chavan}},
  \bibinfo {author} {\bibfnamefont {G.~T.}\ \bibnamefont {Palomino}}, \bibinfo
  {author} {\bibfnamefont {C.~O.}\ \bibnamefont {Are\'{a}n}}, \ and\ \bibinfo
  {author} {\bibfnamefont {S.}~\bibnamefont {Bordiga}},\ }\href@noop {}
  {\bibfield  {journal} {\bibinfo  {journal} {J. Phys. Chem. C}\ }\textbf
  {\bibinfo {volume} {114}},\ \bibinfo {pages} {11185} (\bibinfo {year}
  {2010})}\BibitemShut {NoStop}%
\bibitem [{\citenamefont {Bao}\ \emph {et~al.}(2011)\citenamefont {Bao},
  \citenamefont {Yu}, \citenamefont {Ren}, \citenamefont {Lu},\ and\
  \citenamefont {Deng}}]{bao2011}%
  \BibitemOpen
  \bibfield  {author} {\bibinfo {author} {\bibfnamefont {Z.}~\bibnamefont
  {Bao}}, \bibinfo {author} {\bibfnamefont {L.}~\bibnamefont {Yu}}, \bibinfo
  {author} {\bibfnamefont {Q.}~\bibnamefont {Ren}}, \bibinfo {author}
  {\bibfnamefont {X.}~\bibnamefont {Lu}}, \ and\ \bibinfo {author}
  {\bibfnamefont {S.}~\bibnamefont {Deng}},\ }\href@noop {} {\bibfield
  {journal} {\bibinfo  {journal} {J. Colloid Interface Sci.}\ }\textbf
  {\bibinfo {volume} {353}},\ \bibinfo {pages} {549} (\bibinfo {year}
  {2011})}\BibitemShut {NoStop}%
\bibitem [{\citenamefont {Canepa}\ \emph
  {et~al.}(2013{\natexlab{a}})\citenamefont {Canepa}, \citenamefont {Nijem},
  \citenamefont {Chabal},\ and\ \citenamefont {Thonhauser}}]{piero2013}%
  \BibitemOpen
  \bibfield  {author} {\bibinfo {author} {\bibfnamefont {P.}~\bibnamefont
  {Canepa}}, \bibinfo {author} {\bibfnamefont {N.}~\bibnamefont {Nijem}},
  \bibinfo {author} {\bibfnamefont {Y.~J.}\ \bibnamefont {Chabal}}, \ and\
  \bibinfo {author} {\bibfnamefont {T.}~\bibnamefont {Thonhauser}},\
  }\href@noop {} {\bibfield  {journal} {\bibinfo  {journal} {Phys. Rev. Lett.}\
  }\textbf {\bibinfo {volume} {110}},\ \bibinfo {pages} {026102} (\bibinfo
  {year} {2013}{\natexlab{a}})}\BibitemShut {NoStop}%
\bibitem [{\citenamefont {Canepa}\ \emph
  {et~al.}(2013{\natexlab{b}})\citenamefont {Canepa}, \citenamefont {Arter},
  \citenamefont {Conwill}, \citenamefont {Johnson}, \citenamefont {Shoemaker},
  \citenamefont {Soliman},\ and\ \citenamefont {Thonhauser}}]{piero_01}%
  \BibitemOpen
  \bibfield  {author} {\bibinfo {author} {\bibfnamefont {P.}~\bibnamefont
  {Canepa}}, \bibinfo {author} {\bibfnamefont {C.~A.}\ \bibnamefont {Arter}},
  \bibinfo {author} {\bibfnamefont {E.~M.}\ \bibnamefont {Conwill}}, \bibinfo
  {author} {\bibfnamefont {D.~H.}\ \bibnamefont {Johnson}}, \bibinfo {author}
  {\bibfnamefont {B.~A.}\ \bibnamefont {Shoemaker}}, \bibinfo {author}
  {\bibfnamefont {K.~Z.}\ \bibnamefont {Soliman}}, \ and\ \bibinfo {author}
  {\bibfnamefont {T.}~\bibnamefont {Thonhauser}},\ }\href@noop {} {\bibfield
  {journal} {\bibinfo  {journal} {J. Mater. Chem. A}\ }\textbf {\bibinfo
  {volume} {1}},\ \bibinfo {pages} {13597} (\bibinfo {year}
  {2013}{\natexlab{b}})}\BibitemShut {NoStop}%
  \bibitem [{\citenamefont {Tan}\ \emph {et~al.}(2015)\citenamefont {Tan},
  \citenamefont {Zuluaga}, \citenamefont {Gong}, \citenamefont {Gao},
  \citenamefont {Nijem}, \citenamefont {Li}, \citenamefont {Thonhauser},\ and\
  \citenamefont {Chabal}}]{tan2015competitive}%
  \BibitemOpen
  \bibfield  {author} {\bibinfo {author} {\bibfnamefont {K.}~\bibnamefont
  {Tan}}, \bibinfo {author} {\bibfnamefont {S.}~\bibnamefont {Zuluaga}},
  \bibinfo {author} {\bibfnamefont {Q.}~\bibnamefont {Gong}}, \bibinfo {author}
  {\bibfnamefont {Y.}~\bibnamefont {Gao}}, \bibinfo {author} {\bibfnamefont
  {N.}~\bibnamefont {Nijem}}, \bibinfo {author} {\bibfnamefont
  {J.}~\bibnamefont {Li}}, \bibinfo {author} {\bibfnamefont {T.}~\bibnamefont
  {Thonhauser}}, \ and\ \bibinfo {author} {\bibfnamefont {Y.~J.}\ \bibnamefont
  {Chabal}},\ }\href@noop {} {\bibfield  {journal} {\bibinfo  {journal}
  {Chem. Mater. }\ }\textbf {\bibinfo {volume} {27}},\ \bibinfo
  {pages} {2203} (\bibinfo {year} {2015})}\BibitemShut {NoStop}%
\bibitem [{\citenamefont {Kong}\ \emph {et~al.}(2012)\citenamefont {Kong},
  \citenamefont {Scott}, \citenamefont {Ding}, \citenamefont {Mason},
  \citenamefont {Long},\ and\ \citenamefont {Reimer}}]{long2012}%
  \BibitemOpen
  \bibfield  {author} {\bibinfo {author} {\bibfnamefont {X.}~\bibnamefont
  {Kong}}, \bibinfo {author} {\bibfnamefont {E.}~\bibnamefont {Scott}},
  \bibinfo {author} {\bibfnamefont {W.}~\bibnamefont {Ding}}, \bibinfo {author}
  {\bibfnamefont {J.~A.}\ \bibnamefont {Mason}}, \bibinfo {author}
  {\bibfnamefont {J.~R.}\ \bibnamefont {Long}}, \ and\ \bibinfo {author}
  {\bibfnamefont {J.~A.}\ \bibnamefont {Reimer}},\ }\href@noop {} {\bibfield
  {journal} {\bibinfo  {journal} {J. Am. Chem. Soc.}\ }\textbf {\bibinfo
  {volume} {134}},\ \bibinfo {pages} {14341} (\bibinfo {year}
  {2012})}\BibitemShut {NoStop}%
\bibitem [{\citenamefont {Krishna}\ and\ \citenamefont {van
  Baten}(2012)}]{krishna_jpcc}%
  \BibitemOpen
  \bibfield  {author} {\bibinfo {author} {\bibfnamefont {R.}~\bibnamefont
  {Krishna}}\ and\ \bibinfo {author} {\bibfnamefont {J.~M.}\ \bibnamefont {van
  Baten}},\ }\href@noop {} {\bibfield  {journal} {\bibinfo  {journal} {J. Phys.
  Chem. C}\ }\textbf {\bibinfo {volume} {116}},\ \bibinfo {pages} {23556}
  (\bibinfo {year} {2012})}\BibitemShut {NoStop}%
\bibitem [{\citenamefont {Skoulidas}\ and\ \citenamefont
  {Sholl}(2005)}]{jpcb2005}%
  \BibitemOpen
  \bibfield  {author} {\bibinfo {author} {\bibfnamefont {A.~I.}\ \bibnamefont
  {Skoulidas}}\ and\ \bibinfo {author} {\bibfnamefont {D.~S.}\ \bibnamefont
  {Sholl}},\ }\href@noop {} {\bibfield  {journal} {\bibinfo  {journal} {J.
  Phys. Chem. B}\ }\textbf {\bibinfo {volume} {109}},\ \bibinfo {pages} {15760}
  (\bibinfo {year} {2005})}\BibitemShut {NoStop}%
\bibitem [{\citenamefont {Babarao}\ and\ \citenamefont
  {Jiang}(2008)}]{jiang2008}%
  \BibitemOpen
  \bibfield  {author} {\bibinfo {author} {\bibfnamefont {R.}~\bibnamefont
  {Babarao}}\ and\ \bibinfo {author} {\bibfnamefont {J.}~\bibnamefont
  {Jiang}},\ }\href@noop {} {\bibfield  {journal} {\bibinfo  {journal}
  {Langmuir}\ }\textbf {\bibinfo {volume} {24}},\ \bibinfo {pages} {5474}
  (\bibinfo {year} {2008})}\BibitemShut {NoStop}%
\bibitem [{\citenamefont {Drage}\ \emph {et~al.}(2012)\citenamefont {Drage},
  \citenamefont {Snape}, \citenamefont {Stevens}, \citenamefont {Wood},
  \citenamefont {Wang}, \citenamefont {Cooper}, \citenamefont {Dawson},
  \citenamefont {Guo}, \citenamefont {Satterley},\ and\ \citenamefont
  {Irons}}]{compos}%
  \BibitemOpen
  \bibfield  {author} {\bibinfo {author} {\bibfnamefont {T.~C.}\ \bibnamefont
  {Drage}}, \bibinfo {author} {\bibfnamefont {C.~E.}\ \bibnamefont {Snape}},
  \bibinfo {author} {\bibfnamefont {L.~A.}\ \bibnamefont {Stevens}}, \bibinfo
  {author} {\bibfnamefont {J.}~\bibnamefont {Wood}}, \bibinfo {author}
  {\bibfnamefont {J.}~\bibnamefont {Wang}}, \bibinfo {author} {\bibfnamefont
  {A.~I.}\ \bibnamefont {Cooper}}, \bibinfo {author} {\bibfnamefont
  {R.}~\bibnamefont {Dawson}}, \bibinfo {author} {\bibfnamefont
  {X.}~\bibnamefont {Guo}}, \bibinfo {author} {\bibfnamefont {C.}~\bibnamefont
  {Satterley}}, \ and\ \bibinfo {author} {\bibfnamefont {R.}~\bibnamefont
  {Irons}},\ }\href@noop {} {\bibfield  {journal} {\bibinfo  {journal} {J.
  Mater. Chem.}\ }\textbf {\bibinfo {volume} {22}},\ \bibinfo {pages} {2815}
  (\bibinfo {year} {2012})}\BibitemShut {NoStop}%
   \bibitem [{\citenamefont {Li}\ \emph {et~al.}(2008)\citenamefont {Li},
  \citenamefont {Xiao}, \citenamefont {Webley}, \citenamefont {Zhang}, 
  \citenamefont {Singh}, \ and\ \citenamefont
  {Marshall}}]{gang2008}%
  \BibitemOpen
  \bibfield  {author} {\bibinfo {author} {\bibfnamefont {G.}~\bibnamefont
  {Li}}, \bibinfo {author} {\bibfnamefont {P.}~\bibnamefont {Xiao}},
  \bibinfo {author} {\bibfnamefont {P.}~\bibnamefont {Webley}}, 
  \bibinfo {author} {\bibfnamefont {J.}~\bibnamefont {Zhang}},
  \bibinfo {author} {\bibfnamefont {R.}~\bibnamefont {Singh}},\ and\
  \bibinfo {author} {\bibfnamefont {M.}\ \bibnamefont {Marshall}},\ }\href@noop
  {} {\bibfield  {journal} {\bibinfo  {journal} {Adsorption}\ }\textbf {\bibinfo
  {volume} {14}},\ \bibinfo {pages} {415} (\bibinfo {year}
  {2008})}\BibitemShut {NoStop}%
  \bibitem [{\citenamefont {Kizzie}\ \emph {et~al.}(2011)\citenamefont {Kizzie},
  \citenamefont {Wong-Foy},\ and\ \citenamefont {Matzger}}]{matzger2011}%
  \BibitemOpen
  \bibfield  {author} {\bibinfo {author} {\bibfnamefont {A.~C.}\ \bibnamefont
  {Kizzie}}, \bibinfo {author} {\bibfnamefont {A.~G.}\ \bibnamefont
  {Wong-Foy}}, \ and\ \bibinfo {author} {\bibfnamefont {A.~J.}\ \bibnamefont
  {Matzger}},\ }\href@noop {} {\bibfield  {journal} {\bibinfo  {journal}
  {Langmuir}\ }\textbf {\bibinfo {volume} {27}},\ \bibinfo {pages} {6368}
  (\bibinfo {year} {2011})}\BibitemShut {NoStop}%
\bibitem [{\citenamefont {Nugent}\ \emph {et~al.}(2013)\citenamefont {Nugent},
  \citenamefont {Belmabkhout}, \citenamefont {Burd}, \citenamefont {Cairns},
  \citenamefont {Luebke}, \citenamefont {Forrest}, \citenamefont {Pham},
  \citenamefont {Ma}, \citenamefont {Space}, \citenamefont {Wojtas},
  \citenamefont {Eddaoudi},\ and\ \citenamefont {Zaworotko}}]{patric2013}%
  \BibitemOpen
  \bibfield  {author} {\bibinfo {author} {\bibfnamefont {P.}~\bibnamefont
  {Nugent}}, \bibinfo {author} {\bibfnamefont {Y.}~\bibnamefont {Belmabkhout}},
  \bibinfo {author} {\bibfnamefont {S.~D.}\ \bibnamefont {Burd}}, \bibinfo
  {author} {\bibfnamefont {A.~J.}\ \bibnamefont {Cairns}}, \bibinfo {author}
  {\bibfnamefont {R.}~\bibnamefont {Luebke}}, \bibinfo {author} {\bibfnamefont
  {K.}~\bibnamefont {Forrest}}, \bibinfo {author} {\bibfnamefont
  {T.}~\bibnamefont {Pham}}, \bibinfo {author} {\bibfnamefont {S.}~\bibnamefont
  {Ma}}, \bibinfo {author} {\bibfnamefont {B.}~\bibnamefont {Space}}, \bibinfo
  {author} {\bibfnamefont {L.}~\bibnamefont {Wojtas}}, \bibinfo {author}
  {\bibfnamefont {M.}~\bibnamefont {Eddaoudi}}, \ and\ \bibinfo {author}
  {\bibfnamefont {M.~J.}\ \bibnamefont {Zaworotko}},\ }\href@noop {} {\bibfield
   {journal} {\bibinfo  {journal} {Nature}\ }\textbf {\bibinfo {volume}
  {495}},\ \bibinfo {pages} {80} (\bibinfo {year} {2013})}\BibitemShut
  {NoStop}%
\bibitem [{\citenamefont {Lee}\ \emph {et~al.}(2011)\citenamefont {Lee},
  \citenamefont {Bae}, \citenamefont {Jeong}, \citenamefont {Farha},
  \citenamefont {Sarjeant}, \citenamefont {Stern}, \citenamefont {Nickias},
  \citenamefont {Snurr}, \citenamefont {Hupp},\ and\ \citenamefont
  {Nguyen}}]{yeon2011}%
  \BibitemOpen
  \bibfield  {author} {\bibinfo {author} {\bibfnamefont {C.~Y.}\ \bibnamefont
  {Lee}}, \bibinfo {author} {\bibfnamefont {Y.-S.}\ \bibnamefont {Bae}},
  \bibinfo {author} {\bibfnamefont {N.~C.}\ \bibnamefont {Jeong}}, \bibinfo
  {author} {\bibfnamefont {O.~K.}\ \bibnamefont {Farha}}, \bibinfo {author}
  {\bibfnamefont {A.~A.}\ \bibnamefont {Sarjeant}}, \bibinfo {author}
  {\bibfnamefont {C.~L.}\ \bibnamefont {Stern}}, \bibinfo {author}
  {\bibfnamefont {P.}~\bibnamefont {Nickias}}, \bibinfo {author} {\bibfnamefont
  {R.~Q.}\ \bibnamefont {Snurr}}, \bibinfo {author} {\bibfnamefont {J.~T.}\
  \bibnamefont {Hupp}}, \ and\ \bibinfo {author} {\bibfnamefont {S.~T.}\
  \bibnamefont {Nguyen}},\ }\href@noop {} {\bibfield  {journal} {\bibinfo
  {journal} {J. Am. Chem. Soc.}\ }\textbf {\bibinfo {volume} {133}},\ \bibinfo
  {pages} {5228} (\bibinfo {year} {2011})}\BibitemShut {NoStop}%
\bibitem [{\citenamefont {Li}\ \emph {et~al.}(2009{\natexlab{b}})\citenamefont
  {Li}, \citenamefont {Olson}, \citenamefont {Seidel}, \citenamefont {Emge},
  \citenamefont {Gong}, \citenamefont {Zeng},\ and\ \citenamefont
  {Li}}]{gong2009}%
  \BibitemOpen
  \bibfield  {author} {\bibinfo {author} {\bibfnamefont {K.}~\bibnamefont
  {Li}}, \bibinfo {author} {\bibfnamefont {D.~H.}\ \bibnamefont {Olson}},
  \bibinfo {author} {\bibfnamefont {J.}~\bibnamefont {Seidel}}, \bibinfo
  {author} {\bibfnamefont {T.~J.}\ \bibnamefont {Emge}}, \bibinfo {author}
  {\bibfnamefont {H.}~\bibnamefont {Gong}}, \bibinfo {author} {\bibfnamefont
  {H.}~\bibnamefont {Zeng}}, \ and\ \bibinfo {author} {\bibfnamefont
  {J.}~\bibnamefont {Li}},\ }\href@noop {} {\bibfield  {journal} {\bibinfo
  {journal} {J. Am. Chem. Soc.}\ }\textbf {\bibinfo {volume} {131}},\ \bibinfo
  {pages} {10368} (\bibinfo {year} {2009}{\natexlab{b}})}\BibitemShut {NoStop}%
\bibitem [{\citenamefont {Xue}\ \emph {et~al.}(2015)\citenamefont {Xue},
  \citenamefont {Belmabkhout}, \citenamefont {Shekhah}, \citenamefont {Jiang},
  \citenamefont {Adil}, \citenamefont {Cairns},\ and\ \citenamefont
  {Eddaoudi}}]{king2015}%
  \BibitemOpen
  \bibfield  {author} {\bibinfo {author} {\bibfnamefont {D.-X.}\ \bibnamefont
  {Xue}}, \bibinfo {author} {\bibfnamefont {Y.}~\bibnamefont {Belmabkhout}},
  \bibinfo {author} {\bibfnamefont {O.}~\bibnamefont {Shekhah}}, \bibinfo
  {author} {\bibfnamefont {H.}~\bibnamefont {Jiang}}, \bibinfo {author}
  {\bibfnamefont {K.}~\bibnamefont {Adil}}, \bibinfo {author} {\bibfnamefont
  {A.~J.}\ \bibnamefont {Cairns}}, \ and\ \bibinfo {author} {\bibfnamefont
  {M.}~\bibnamefont {Eddaoudi}},\ }\href@noop {} {\bibfield  {journal}
  {\bibinfo  {journal} {J. Am. Chem. Soc.}\ }\textbf {\bibinfo {volume}
  {137}},\ \bibinfo {pages} {5034} (\bibinfo {year} {2015})}\BibitemShut
  {NoStop}%
\bibitem [{\citenamefont {Liu}\ \emph {et~al.}(2012{\natexlab{b}})\citenamefont
  {Liu}, \citenamefont {Tian}, \citenamefont {Thallapally},\ and\ \citenamefont
  {McGrail}}]{jian2012}%
  \BibitemOpen
  \bibfield  {author} {\bibinfo {author} {\bibfnamefont {J.}~\bibnamefont
  {Liu}}, \bibinfo {author} {\bibfnamefont {J.}~\bibnamefont {Tian}}, \bibinfo
  {author} {\bibfnamefont {P.~K.}\ \bibnamefont {Thallapally}}, \ and\ \bibinfo
  {author} {\bibfnamefont {B.~P.}\ \bibnamefont {McGrail}},\ }\href@noop {}
  {\bibfield  {journal} {\bibinfo  {journal} {J. Phys. Chem. C}\ }\textbf {\bibinfo
  {volume} {116}},\ \bibinfo {pages} {9575} (\bibinfo {year}
  {2012}{\natexlab{b}})}\BibitemShut {NoStop}%
\bibitem [{Note1()}]{Note1}%
  \BibitemOpen
  \bibinfo {note} {In Fig.~\ref {fig:fig_SI02} we show that our qualitative
  conclusions are unchanged if we take the relative abundance of gas types to
  be typical of a flue-gas mixture.}\BibitemShut {Stop}%
  \bibitem [{\citenamefont {Poloni}\ \emph {et~al.}(2008)\citenamefont {Poloni},
  \citenamefont {Lee}, \citenamefont {Berger}, \citenamefont {Smit},
  \ and\ \citenamefont {Neaton}}]{poloni_jpcl}%
  \BibitemOpen
  \bibfield  {author} {\bibinfo {author} {\bibfnamefont {R.}~\bibnamefont
  {Poloni}}, \bibinfo {author} {\bibfnamefont {K.}~\bibnamefont {Lee}}, \bibinfo
  {author} {\bibfnamefont {R.~F.}~\bibnamefont {Berger}}, \bibinfo {author} 
  {\bibfnamefont {B.}~\bibnamefont {Smit}}, \ and\ \bibinfo {author} {\bibfnamefont
  {J.~B.}~\bibnamefont {Neaton}},\ }\href@noop {} {\bibfield  {journal} {\bibinfo
  {journal} {J. Phys. Chem. Lett.}\ }\textbf {\bibinfo {volume} {5}},\ \bibinfo
  {pages} {861} (\bibinfo {year} {2014})}\BibitemShut {NoStop}%
\bibitem [{\citenamefont {Liu}\ \emph {et~al.}(2008)\citenamefont {Liu},
  \citenamefont {Yang}, \citenamefont {Xue}, \citenamefont {Zhong},
  \citenamefont {Chen},\ and\ \citenamefont {Smit}}]{liu2008}%
  \BibitemOpen
  \bibfield  {author} {\bibinfo {author} {\bibfnamefont {B.}~\bibnamefont
  {Liu}}, \bibinfo {author} {\bibfnamefont {Q.}~\bibnamefont {Yang}}, \bibinfo
  {author} {\bibfnamefont {C.}~\bibnamefont {Xue}}, \bibinfo {author}
  {\bibfnamefont {C.}~\bibnamefont {Zhong}}, \bibinfo {author} {\bibfnamefont
  {B.}~\bibnamefont {Chen}}, \ and\ \bibinfo {author} {\bibfnamefont
  {B.}~\bibnamefont {Smit}},\ }\href@noop {} {\bibfield  {journal} {\bibinfo
  {journal} {J. Phys. Chem. C}\ }\textbf {\bibinfo {volume} {112}},\ \bibinfo
  {pages} {9854} (\bibinfo {year} {2008})}\BibitemShut {NoStop}%
\bibitem [{\citenamefont {Liu}\ and\ \citenamefont {Smit}(2009)}]{liu2009}%
  \BibitemOpen
  \bibfield  {author} {\bibinfo {author} {\bibfnamefont {B.}~\bibnamefont
  {Liu}}\ and\ \bibinfo {author} {\bibfnamefont {B.}~\bibnamefont {Smit}},\
  }\href@noop {} {\bibfield  {journal} {\bibinfo  {journal} {Langmuir}\
  }\textbf {\bibinfo {volume} {25}},\ \bibinfo {pages} {5918} (\bibinfo {year}
  {2009})}\BibitemShut {NoStop}%
\bibitem [{\citenamefont {Braun}\ \emph {et~al.}(2015)\citenamefont {Braun},
  \citenamefont {Chen}, \citenamefont {Schnell}, \citenamefont {Lin},
  \citenamefont {Reimer},\ and\ \citenamefont {Smit}}]{barun2015}%
  \BibitemOpen
  \bibfield  {author} {\bibinfo {author} {\bibfnamefont {E.}~\bibnamefont
  {Braun}}, \bibinfo {author} {\bibfnamefont {J.~J.}\ \bibnamefont {Chen}},
  \bibinfo {author} {\bibfnamefont {S.~K.}\ \bibnamefont {Schnell}}, \bibinfo
  {author} {\bibfnamefont {L.-C.}\ \bibnamefont {Lin}}, \bibinfo {author}
  {\bibfnamefont {J.~A.}\ \bibnamefont {Reimer}}, \ and\ \bibinfo {author}
  {\bibfnamefont {B.}~\bibnamefont {Smit}},\ }\href@noop {} {\bibfield
  {journal} {\bibinfo  {journal} {Angew. Chem. Int. Ed.}\ }\textbf {\bibinfo
  {volume} {54}},\ \bibinfo {pages} {14349} (\bibinfo {year}
  {2015})}\BibitemShut {NoStop}%
\bibitem [{\citenamefont {Lopez}\ \emph {et~al.}(2013)\citenamefont {Lopez},
  \citenamefont {Canepa},\ and\ \citenamefont {Thonhauser}}]{piero_02}%
  \BibitemOpen
  \bibfield  {author} {\bibinfo {author} {\bibfnamefont {M.~G.}\ \bibnamefont
  {Lopez}}, \bibinfo {author} {\bibfnamefont {P.}~\bibnamefont {Canepa}}, \
  and\ \bibinfo {author} {\bibfnamefont {T.}~\bibnamefont {Thonhauser}},\
  }\href@noop {} {\bibfield  {journal} {\bibinfo  {journal} {J. Chem. Phys.}\
  }\textbf {\bibinfo {volume} {138}},\ \bibinfo {pages} {154704} (\bibinfo
  {year} {2013})}\BibitemShut {NoStop}%
\bibitem [{\citenamefont {Queen}\ \emph {et~al.}(2011)\citenamefont {Queen},
  \citenamefont {Brown}, \citenamefont {Britt}, \citenamefont {Zajdel},
  \citenamefont {Hudson},\ and\ \citenamefont {Yaghi}}]{queen2011}%
  \BibitemOpen
  \bibfield  {author} {\bibinfo {author} {\bibfnamefont {W.~L.}\ \bibnamefont
  {Queen}}, \bibinfo {author} {\bibfnamefont {C.~M.}\ \bibnamefont {Brown}},
  \bibinfo {author} {\bibfnamefont {D.~K.}\ \bibnamefont {Britt}}, \bibinfo
  {author} {\bibfnamefont {P.}~\bibnamefont {Zajdel}}, \bibinfo {author}
  {\bibfnamefont {M.~R.}\ \bibnamefont {Hudson}}, \ and\ \bibinfo {author}
  {\bibfnamefont {O.~M.}\ \bibnamefont {Yaghi}},\ }\href@noop {} {\bibfield
  {journal} {\bibinfo  {journal} {J. Phys. Chem. C}\ }\textbf {\bibinfo
  {volume} {115}},\ \bibinfo {pages} {24915} (\bibinfo {year}
  {2011})}\BibitemShut {NoStop}%
\end{thebibliography}

%

\onecolumngrid
\clearpage

\renewcommand{\theequation}{S\arabic{equation}}
\renewcommand{\thefigure}{S\arabic{figure}}
\renewcommand{\thesection}{S\arabic{section}}

\setcounter{equation}{0}
\setcounter{section}{0}
\setcounter{figure}{0}

\setlength{\parskip}{0.25cm}%
\setlength{\parindent}{0pt}%

\section{supplementary Information for ``Selective gas capture via kinetic trapping"}

\section{Model dynamics}
\label{section_s1}

We evolved the lattice model (\f{fig:fig_SI01}) using a semi-grand canonical Monte Carlo algorithm that allowed single-particle insertion, removal, and diffusion processes. We took the basic microscopic rates for insertion, removal and diffusion processes to be $R_{{\rm i}}$, $R_{{\rm r}}$, and $R_{\rm d}$, respectively. We define the `total rate' $R \equiv L_y R_{{\rm i}} + n_0 R_{{\rm r}} + n R_{{\rm d}}$, where $n_0$ is the instantaneous number of particles on the first column of the lattice, and $n$ is the instantaneous total number of particles on the lattice. With respective probabilities $L_y R_{{\rm i}}/R$, $n_0 R_{{\rm r}}/R$, or $n R_{{\rm d}}/R$ we attempted an insertion, a deletion, or a diffusion move. To attempt an insertion we chose uniformly one of the $L_y$ sites in the first column of the lattice, and attempted to place a particle on that site. The particle was chosen to be of type corresponding to \h, \cot, or \water~with equal likelihood. To attempt a removal we chose uniformly any of the $n_0$ particles on the first column of the lattice, and proposed to remove it from the lattice. To attempt a diffusion move we chose uniformly one of the $n$ particles on the lattice, and proposed with uniform likelihood to move that particle to any one of its four nearest-neighbor sites. We accepted each proposed move with probability
\be
\label{rate}
p_{\rm acc} = \min \left(1,\frac{R_{\rm before}}{R_{\rm after}} {\rm e}^{-\beta \Delta E} \right).
\ee
Here $R_{\rm before}$ and $R_{\rm after}$ are the values of the total rate $R$ before and after the proposed move; $\beta \equiv 1/(k_{\rm B}T)$; and $\Delta E$ is the energy change resulting from the proposed move. This energy change accounts for particle-framework binding energies, hard-core particle site exclusions, and particle-particle nearest-neighbor interactions. For diffusion moves, in addition, any proposal to take a particle across the left or right extremity of the simulation box was rejected. For insertion and removal moves the form (\ref{rate}) corresponds to the choice of fixed chemical potential $\mu = - k_{\rm B} T \ln (3 R_{\rm r}/R_{\rm i})$. For the simulations presented in the paper we set  $R_{\rm r}=2$ and $R_{\rm i}=5$. After every proposed move we updated time by an amount $(R_{\rm d}/R) \Delta t$, where $\Delta t = 10^{-10}$ s is the basic timescale of our model (see the next section for details).

The results presented in the main text are for equimolar gas mixtures, but we have also carried out simulations for gas mixtures of variable composition. The composition of a post-combustion flue gas 
depends on the particulars of the power plant. Often it consists of 70-75\% N$_2$, 12-15\% 
CO$_2$, 5-7\% H$_2$O, and small amounts of SO$_2$, NO$_x$, O$_2$, CO, etc. H$_2$ is 
an important constituent of pre-combustion flue gas, and for Mg-MOF-74 occupies the same position as N$_2$ in the hierarchy of binding enthalpies and diffusion barriers with respect to CO$_2$ and H$_2$O.
In \f{fig:fig_SI02}(c) we show a time-temperature plot for the density of CO$_2$ when the model framework is exposed to a gas mixture with composition $75\%$ \h, $15\%$ CO$_2$ and 
$10\%$ H$_2$O, similar to that of a flue gas. For these simulations the chemical potential or fugacities of the three gas types are different and chosen such that $z_{\rm{H_2}}$:$z_{\rm{CO_2}}$:$z_{\rm{H_2O}}$=75:15:10, where $z_{j}=e^{\mu_j /k_{\rm B} T}$ is the fugacity of a particle corresponding to gas type $j$. Algorithmically we proceed in the same way as described above, with a modified insertion move. We attempt insertion moves with probability $L_y R_{\rm{i}}/R$, where now the overall rate of insertion is $R_{\rm{i}}=R_{\rm H_2}+R_{\rm CO_2}+R_{\rm H_2O}$, with $R_{\rm H2}$, $R_{\rm CO_2}$, and $R_{\rm H_2O}$ being the rates of insertion of particles of type $\rm{H_2}$, $\rm{CO_2}$, and $\rm{H_2O}$, respectively. We choose individual rates so that
\be
z_{\rm{H_2}}+z_{\rm{CO_2}}+z_{\rm{H_2O}}=\frac{1}{R_{\rm r}}(R_{\rm H_2}+R_{\rm CO_2}+R_{\rm H_2O}),
\ee
where  $R_{\rm r}$ is the removal rate. Upon choosing an insertion move we propose to insert a particle of gas type $j$ with probability $R_j/R_{\rm{i}}$. We set $R_{\rm{H_2}}=4.5$, $R_{\rm{CO_2}}=0.9$, $R_{\rm{H_2O}}=0.6$, and $R_{\rm r}=1$ so that $z_{\rm{H_2}}$:$z_{\rm{CO_2}}$:$z_{\rm{H_2O}}$=75:15:10. The results of these simulations are qualitatively similar to those of the equimolar simulations, but with \cot~captured in slightly greater abundance.

\section{Estimate of the basic time scale $\Delta t$ of the model}
\label{section_s2}

We determined the basic timescale of our model by comparison with experiment as follows. We set  up a simulation to mimic an experiment in which \water~is passed through Mg-MOF-74 preloaded with bound \cot\c{tan2015competitive}. We measured the time taken by \water~to displace \cot. We started the simulation with all binding sites of the framework occupied by \cot, and then exposed the framework to \water. As time progresses, \water~displaces \cot, as shown in \f{fig:fig_SI03}(a). We define the equilibration time as the time when the bound fraction of \water~is 99\%. \f{fig:fig_SI03}(b) shows the equilibration time $\tau_{\rm eq}$ as a function of $L_z$ at different temperatures. These data can be fit by the equation
\beq
\tau_{\rm eq} (T)=\mbox{A}(T)+\mbox{B}(T)~L^{\nu (T)}_z.
\label{eq:SI02}
\eeq
We summarize the values of A, B and $\nu$ at different temperatures, obtained by fitting the simulation data with Eq.~\ref{eq:SI02}, in table~\ref{table}. We find that $\nu \approx 2$. 
\begin{table}[h]
\begin{tabular}{ |c|c|c|c|c} 
\hline
T (K) & A & B & $\nu$\\
\hline
  523 & $1.26 \times 10^6$ & $8.87$  & $2.05$ \\ 
\hline
  493 & $2.53\times 10^6 $ & $15.49$  & $1.97$ \\ 
\hline
  463 & $5.57 \times 10^6$ & $17.94$  & $1.96$ \\ 
\hline
 433 & $1.46\times 10^{7}$ & $10.02$  & $2.05$ \\ 
\hline
 403 & $4.41\times 10^7$ & $2.21$  & $2.26$ \\ 
\hline
 348 & $5.15 \times 10^8$ & & \\
\hline
\end{tabular}
\caption{Values of A, B and $\nu$ at different temperatures, measured by fitting the simulation data with Eq.~\ref{eq:SI02}.}
\label{table}
\end{table}
In the experiment with a sample of thickness $40~\mu$m, \water~adsorption was found to reach equilibrium after $\sim 210$ s at $348~\mbox{K}$ \c{tan2015competitive}. Such lengthscales and timescales are out of reach of our simulations, but by extrapolation we can compare those experiments with our results. At $348~\mbox{K}$ we can estimate the value of $\mbox{A} (\approx 5.15 \times 10^8)$ by extrapolation. Using \eqq{eq:SI02} and all combinations of B and $\nu$ for the different temperatures listed in table~\ref{table}, we find that the choice $\Delta t \sim 10^{-10}~\mbox{s}$ gives the range $\tau_{\rm eq} (348~\mbox{K})\approx 76-405~\mbox{s}$, which encompasses the experimental value of 210 s. The value of $\Delta t$ is physically reasonable: it is close to the measured self-diffusion time of \cot~in Mg-MOF-74, which is $\sim 10^{-10}$--$10^{-9}~\mbox{s}$~\c{bao2011}. By comparison, the the self-diffusion time of \cot~in air is $\sim 10^{-16}$--$10^{-15}~\mbox{s}$.

\begin{figure}
\includegraphics[width=0.65\columnwidth]{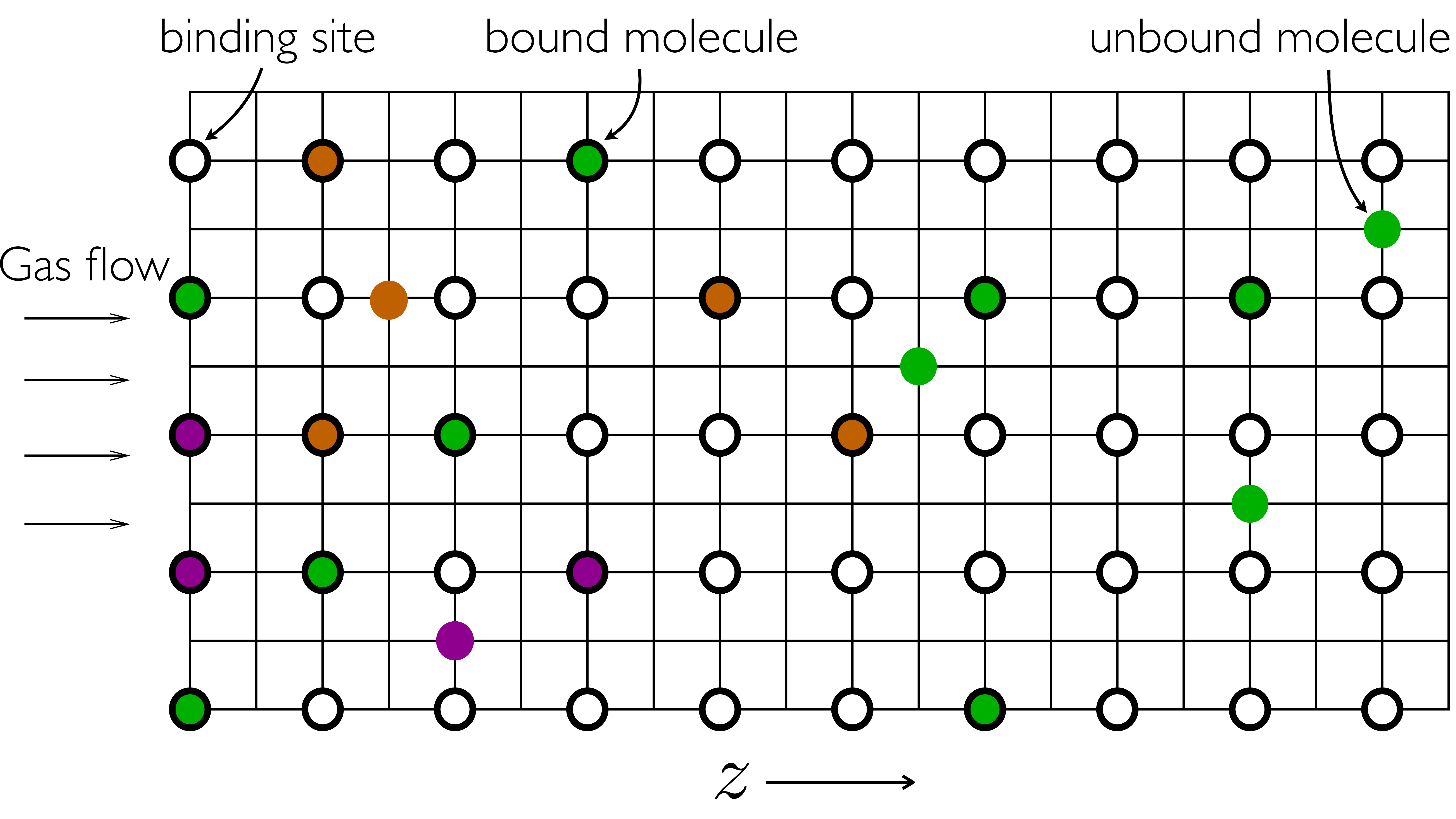}
\caption{Lattice model of a framework for gas capture. The leftmost column ($z=0$) is in contact with a gas reservoir. Sites denoted by bold circles are binding sites. Gas molecules on binding sites receive a favorable energetic interaction. Different colors correspond to different gas types. The alternating rows running along $z$-axis with no binding sites correspond to the one-dimensional channels of a real framework along $c$-axis. In this schematic, the length of the framework along $z$ axis ($L_z$) is 20 lattice sites or equivalently 30 \AA.  Typical size of a single crystal in an experiment is $5-25~\mu \mbox{m}$\c{bao2011}.}
\label{fig:fig_SI01}
\end{figure}

\begin{figure}
\includegraphics[width=\columnwidth]{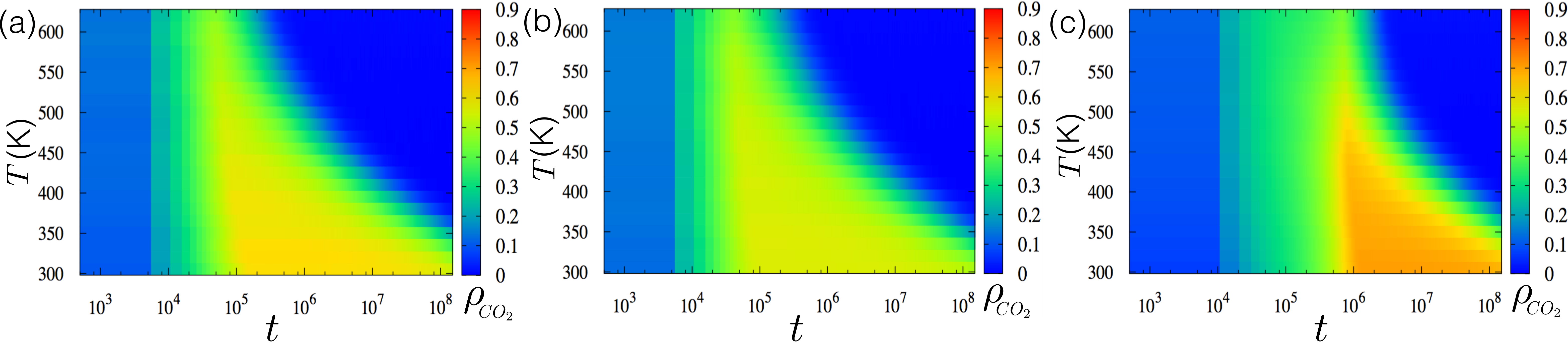}
\caption{Bound \cot~fraction as a function of time $t$, for simulations run at a range of temperature $T$. In panel (a), the values of intra-species pairwise nearest-neighbor repulsive interactions are 0.0025~eV, 0.02~eV, or 0.03~eV for \h, \cot~or \water, respectively\c{piero2013}. Inter-species interactions are chosen to be the arithmetic mean of the corresponding intra-species interactions. In panel (b), inter-species interactions are set to zero. In both the cases the framework was in contact with a reservoir of an equimolar gas mixture. In panel (c), the reservoir contains a mixture of $75 \%$ H$_2$, $15\%$ CO$_2$ and $10\%$ H$_2$O. Here, the inter-species interactions are the same as those used in panel (a). The data are for $L_y=40$ and $L_z=80$.}
\label{fig:fig_SI02}
\end{figure}

\begin{figure}
\includegraphics[width=0.9\columnwidth]{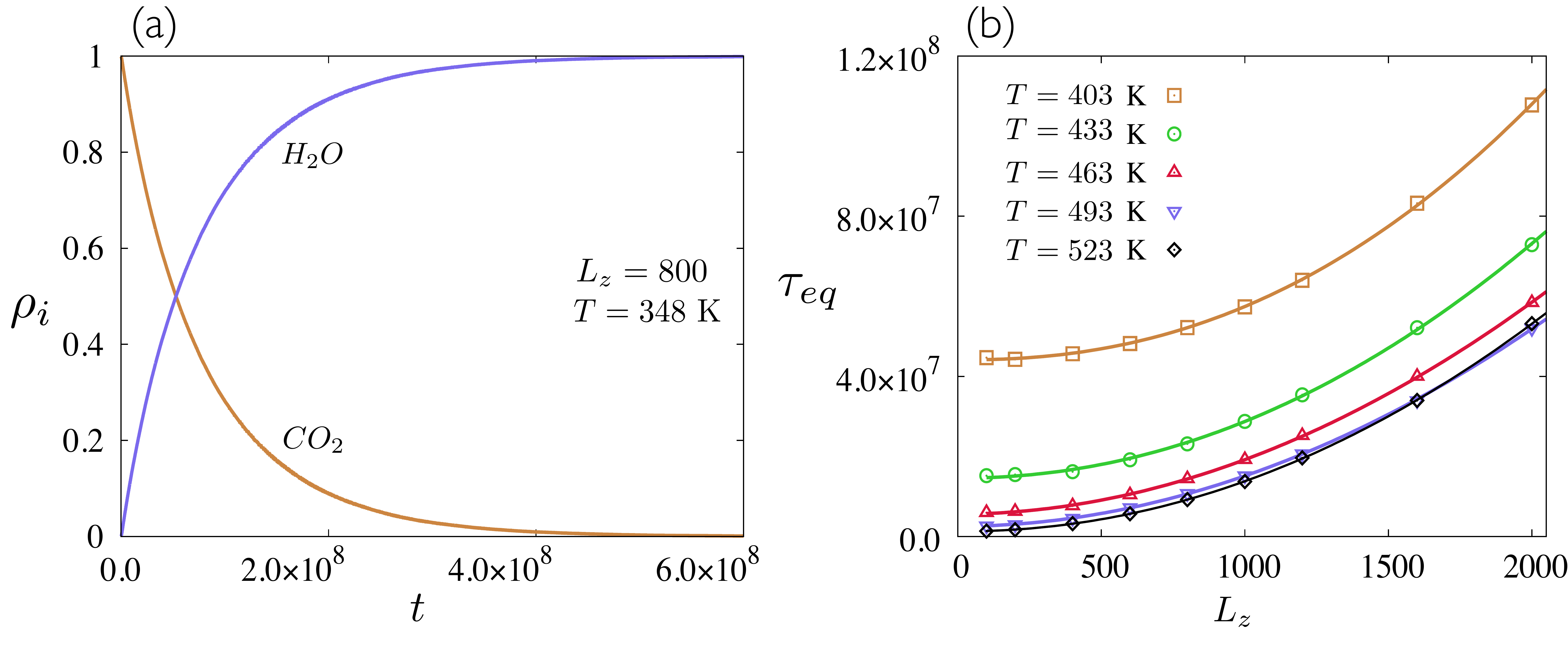}
\caption{We fix the basic time scale $\Delta t$ of our model by comparing the equilibration time in simulation with that measured in an experiment in which a framework pre-loaded with \cot~is exposed to \water\c{tan2015competitive}. (a) Time evolution of the bound fraction of \cot~ and \water~at 348 K for $L_z=800$. (b) Equilibration time $\tau_{\rm eq}$ as a function of $L_z$ at different temperatures. Data points are obtained from simulations. Solid lines are fits to the data using $\tau_{\rm eq} (T)=A(T)+B(T)~L^{\nu (T)}_z$. The values of $A(T)$, $B(T)$, and $\nu(T)$ at different temperatures are listed in table~\ref{table}. The data are averaged over 100 independent simulations and are for $L_y=10$.}
\label{fig:fig_SI03}
\end{figure}

\begin{figure}
\includegraphics[width=0.65\columnwidth]{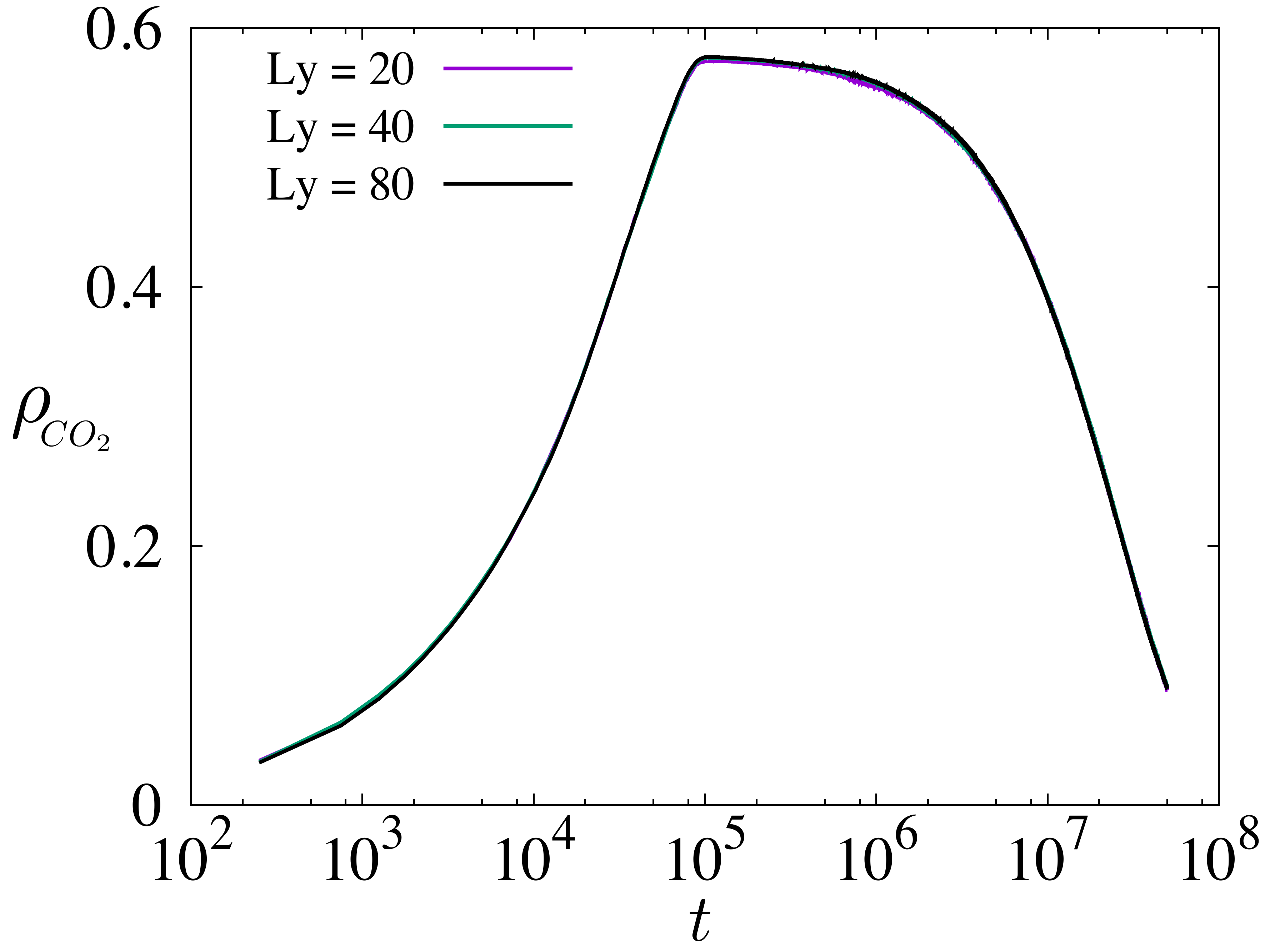}
\caption{Simulation results are largely insensitive to the lateral extent $L_y$ of the framework, because motion along channels is effectively one dimensional. We plot $\rho_{{\rm CO}_{\rm 2}}$ as a function of time $t$ (in units of $\Delta t$) for three different values of $L_y$ ($L_z=80$) at $388$ K, averaged over 100 independent runs.}
\label{fig:fig_SI04}
\end{figure}

\section{Polydispersity of grain size and nonequilibrium gas capture }
\label{section_s3}
Nonequilibrium gas capture can occur simultaneously, in our model, for systems having a range of values of $L_z$, indicating that nonequilibrium gas capture can be effected in a MOF powder whose grains possess a distribution of sizes. In \f{fig:fig_SI05} we plot the bound fraction of \cot~as a function of $L_z$, at different times and temperatures. This plot indicates that for certain conditions an appreciable fraction of \cot~can be bound within grains of a range of sizes (assuming that all grains are exposed to gas almost at the same time, as would occur in a thin bed over which gas is flowed).

 \f{fig:fig_SI06}(a) shows the bound fraction of \cot~($\rho_{CO_2}$) as a function of time for different values of $L_z$ at $328~\mbox{K}$. The data for the time $\tau_{0.5}$ at which $\rho_{{\rm CO}_{\rm 2}}$ decays to a value of 0.5, after attaining its maximum value, are plotted against $L_z$ in \f{fig:fig_SI06}(b). These data can be fit, at 328 K, by $\tau_{0.5}=C+D~L^{m}_z$, where $C\approx7.45\times 10^7$, $D\approx 87.71$ and $m \approx 1.77$. Using this fit we can predict the time window within which \cot~can be captured by a framework. Our results suggest that at $298~\mbox{K}$, a framework of length $L_z = L_c\approx 0.43~\mu \mbox{m}$, exposed to an equimolar mixture of the three gas types considered here, will harbor \cot~at more than 40\% of its binding sites up to a time of $\sim 0.1$ s. Here, $C(298~\mbox{K})\approx 1.15 \times 10^9$, as obtained from simulations at $298~\mbox{K}$ and the values $D$, $m$ are considered to be same as those at $328\mbox{K}$. For a fixed $L_z$, the time window within which \cot~can be captured can be enlarged either by decreasing the temperature or by impeding the gas invasion into the framework (see \f{fig:fig_SI07}).  
\begin{figure}
\includegraphics[width=0.6\columnwidth]{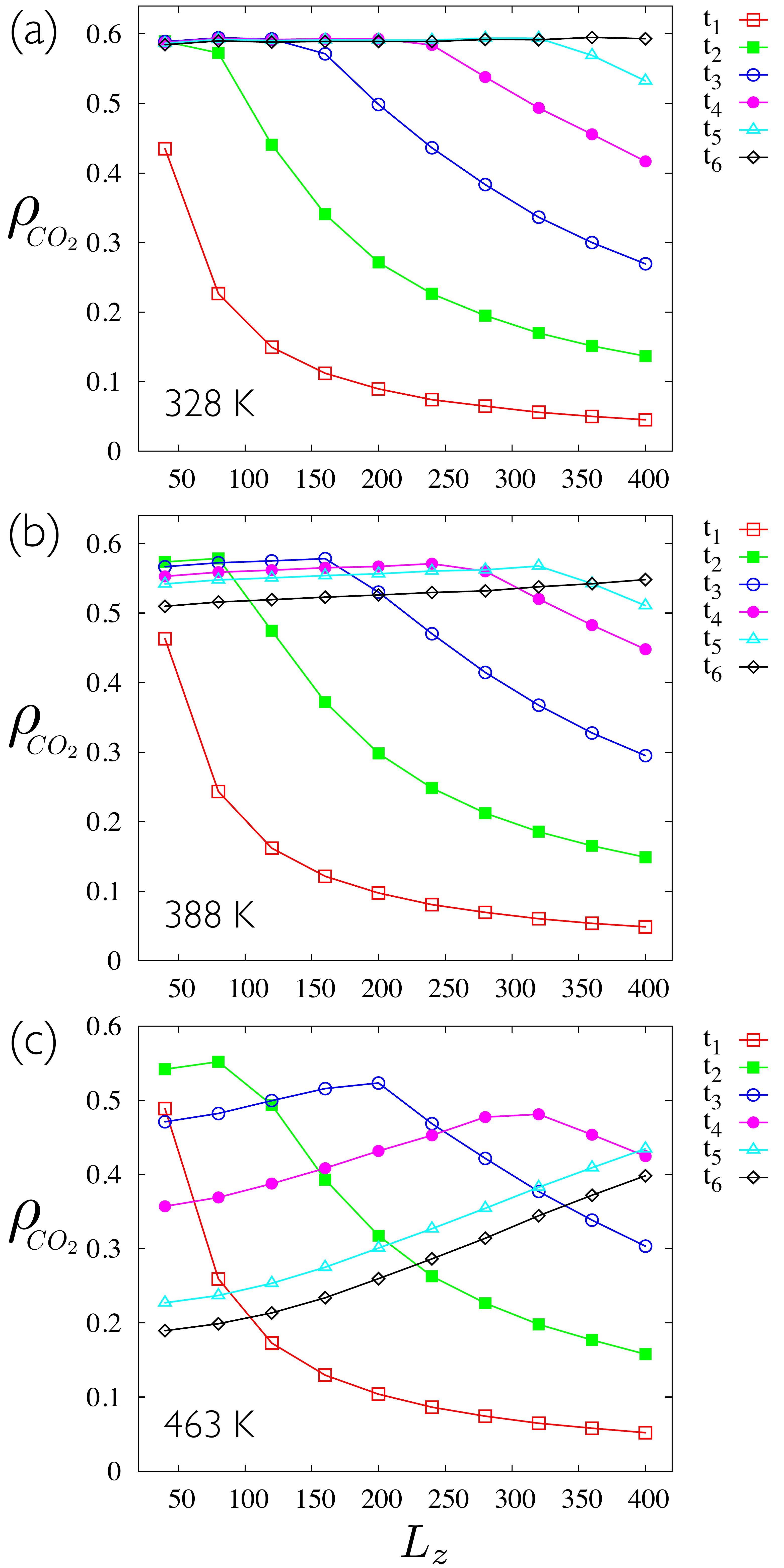}
\caption{The total amount of bound \cot~under nonequilibrium conditions at a given time depends on $L_z$. We show the fraction of binding sites occupied by \cot~as a function of $L_z$ at six different times, for $T=$ (a) $328$ K, (b) $388$ K and (c) $463$ K. Here $t_1<t_2<t_3<t_4<t_5<t_6$. The data, averaged over $200$ independent runs, are for $L_y=40$. At any temperature $T$, $t_1$ and $t_6$ represent times before and after $\rho_{{\rm CO}_{\rm 2}}$ reaches its maximum respectively for any $L_z$. For all the three temperatures, $t_1\approx 10^{4}$. $t_6 \approx 3.5 \times10^6$, $3.0 \times 10^6$ and $2.4\times 10^4$ when $T= 328$ K, $388$ K and $463$ K respectively. To maximize \cot~capture per unit volume of a MOF-bed at a particular time, one should consider a size distribution of grains peaked at the value of $L_z$ where density of \cot~is maximum for that time and temperature. }
\label{fig:fig_SI05}
\end{figure}

\begin{figure}
\includegraphics[width=0.95\columnwidth]{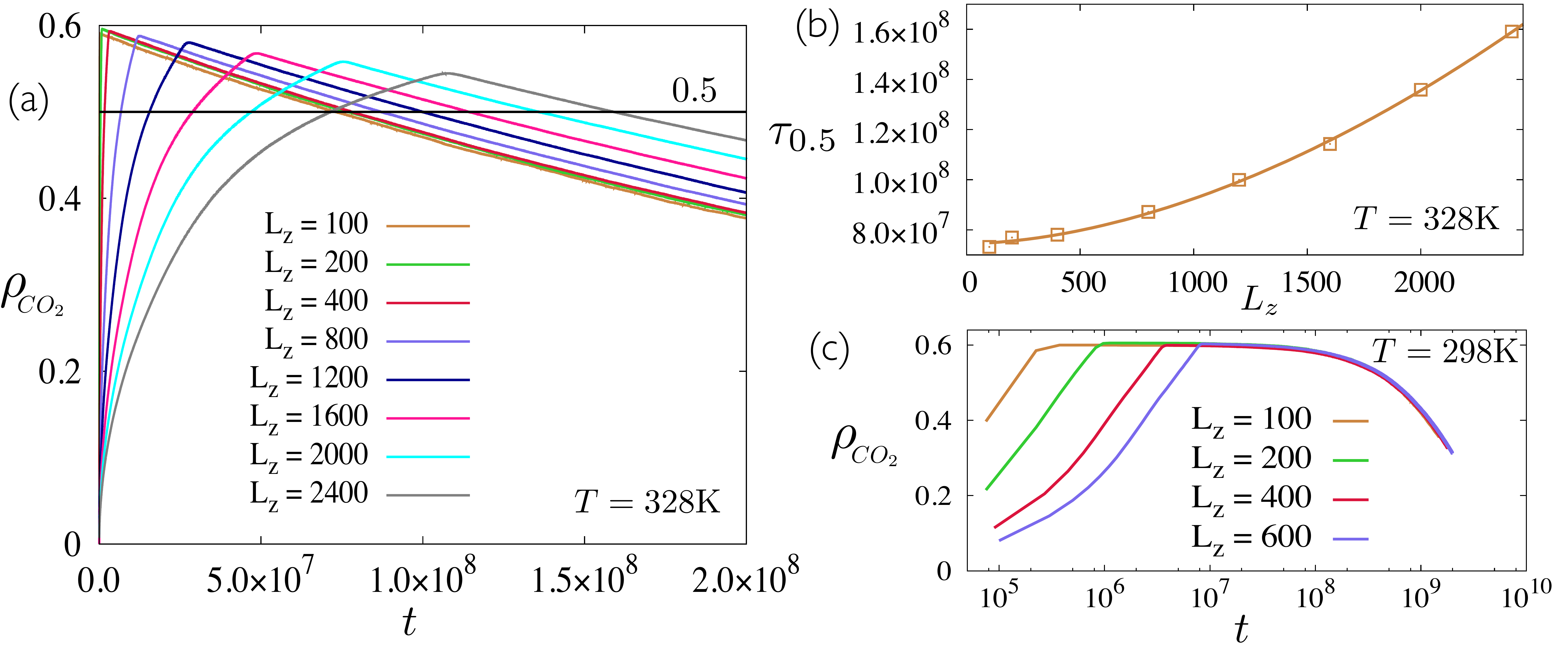}
\caption{We can predict the timescale of \cot~capture, for fixed $L_z$, under nonequilibrium conditions. (a) Time evolution of $\rho_{{\rm CO}_{\rm 2}}$, the bound fraction of \cot~at 328 K for different $L_z$. The horizontal line corresponds to $\rho_{{\rm CO}_{\rm 2}}=0.5$. (b) The time at which $\rho_{{\rm CO}_{\rm 2}}$ decays to 0.5 as a function of the linear extent of the framework along $z$ at 328 K. The data points are obtained from simulations; the solid line is a fit to the data and follows the equation $\tau_{0.5}=C+D~L^{m}_z$, where $C\approx7.45\times 10^7$, $D\approx 87.71$ and $m\approx 1.77$. (c) Time evolution of $\rho_{{\rm CO}_{\rm 2}}$ at 298 K for four different system sizes. We find $C (298 ~\mbox{K}) \approx 1.15 \times 10^9$ by extrapolation. Being unable to go up to larger system sizes, we consider that the value of $D$ and $m$ at 298 K are same as those for 328 K. The data are averaged over 80 independent simulations and are for $L_y=10$.}
\label{fig:fig_SI06}
\end{figure}

\begin{figure}
\includegraphics[width=0.65\columnwidth]{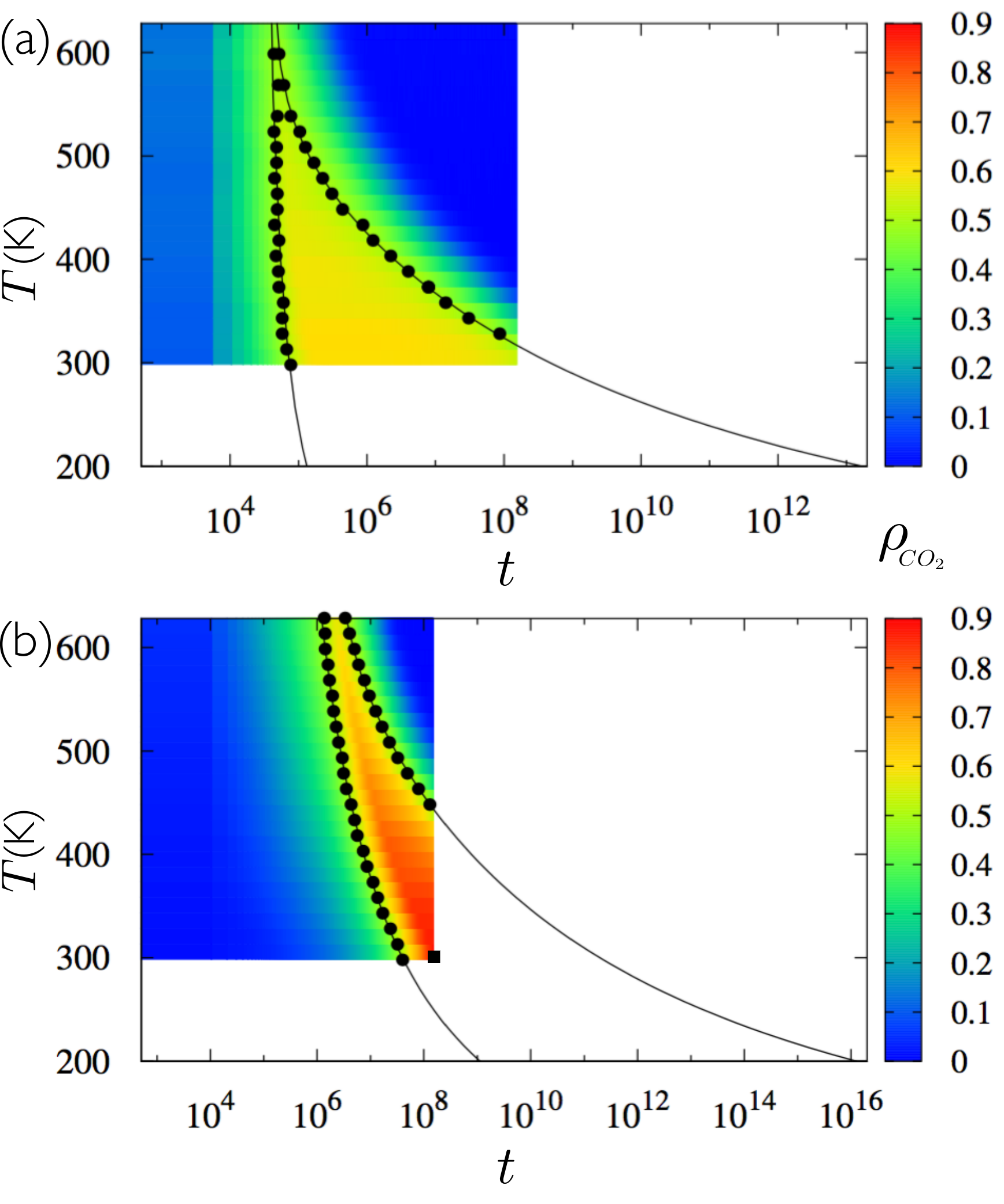}
\caption{The nonequilibrium \cot~capture window can be shifted and broadened by increasing the separation of timescales of basic microscopic processes. We plot the bound \cot~fraction, as a function of $t$, for simulations run at a range of temperature $T$. In panel (a) the values of the barriers opposing the diffusion in a crowded framework of \h, CO$_2$ and H$_2$O molecules are taken from quantum mechanical simulations\c{piero2013}. In panel (b) we have increased all barriers by an arbitrary factor of 5, in order to demonstrate that impeding the flow of all gases into the framework can increase the residence time and maximum abundance of a desired gas. Colored data points are obtained by averaging over $10$ independent simulations (box dimensions $40 \times 80$). Circles corresponds to the points at which the framework is half-full of \cot, i.e. where $\rho_{{\rm CO_2}} = 0.5$. At the point shown by the square in (b), $\rho_{{\rm CO_2}} \approx 0.9$. Curves are Arrhenius fits to the data, and have been extrapolated to lower temperature.}
\label{fig:fig_SI07}
\end{figure}

\end{document}